\documentclass{aa}
\usepackage{natbib}
\usepackage{graphicx}
\usepackage{placeins}
\usepackage{txfonts}
\usepackage{hyperref}

\begin{document}

\newcommand{\kms}{km~s$^{-1}$}
\newcommand{\ms}{m~s$^{-1}$}
\newcommand{\teff}{$T_{\rm eff}$}
\newcommand{\logg}{log~$g$}
\newcommand{\mh}{[M/H]}
\newcommand{\feh}{[Fe/H]}
\newcommand{\am}{[$\alpha$/M]}
\newcommand{\cm}{[C/M]}
\newcommand{\nm}{[N/M]}

\newcommand{\om}{[O/M]}
\newcommand{\mgm}{[Mg/M]}  
\newcommand{\alm}{[Al/M]} 
\newcommand{\simm}{[Si/M]} 
\newcommand{\pmm}{[P/M]} 
\newcommand{\sm}{[S/M]}  
\newcommand{\km}{[K/M]}  
\newcommand{\cam}{[Ca/M]}  
\newcommand{\tim}{[Ti/M]} 
\newcommand{\vm}{[V/M]} 
\newcommand{\crm}{[Cr/M]}  
\newcommand{\mnm}{[Mn/M]}  
\newcommand{\com}{[Co/M]}  
\newcommand{\nim}{[Ni/M]}  
\newcommand{\cum}{[Cu/M]}  
\newcommand{\cem}{[Ce/M]}  
\newcommand{\ndm}{[Nd/M]}  

\newcommand{\snr}{S/N}
\newcommand{\vk}{V$-$K$_{\rm s}$}
\newcommand{\jk}{J$-$K$_{\rm s}$}
\newcommand{\bprp}{BP$-$RP}
\newcommand{\vmicro}{$v_{\rm micro}$}
\newcommand{\vmacro}{$v_{\rm macro}$}
\newcommand{\vrad}{$v_{\rm rad}$}
\newcommand{\ebv}{$E(B-V)$}
\newcommand{\vsini}{$v \ \mathrm{sin} \ i$}
\newcommand{\mum}{$\mu$m}

\title{Radial velocity and atmospheric parameter calculations for the GaiaNIR spectrograph}

\author{
Szabolcs~M{\'e}sz{\'a}ros\inst{1,2,3}, 
David~Hobbs\inst{4}, 
Anna~Liptrott\inst{5},
David~Katz\inst{6},
Ricardo~Schiavon\inst{5},
Vikt{\'o}ria~Pap\inst{1,2},
Anthony~G.~A.~Brown\inst{7},
George Seabroke\inst{8}, 
Joss~Bland-Hawthorn\inst{9}, 
Ronny~Blomme\inst{10}, 
Nicholas~A.~Walton\inst{11}
}

\institute{
ELTE E\"otv\"os Lor\'and University, Gothard Astrophysical Observatory, 9700 Szombathely, Szent Imre H. st. 112, Hungary
\and
HUN-REN CSFK, Konkoly Observatory, Konkoly Thege Mikl\'os \'ut 15-17, Budapest, 1121, Hungary
\and
MTA-ELTE Lend{\"u}let "Momentum" Milky Way Research Group, Hungary
\and
Lund Observatory, Division of Astrophysics, Department of Physics, Lund University, Box 118, SE-22100, Lund, Sweden
\and
Astrophysics Research Institute, Liverpool John Moores University, 146 Brownlow Hill, Liverpool L3 5RF, UK
\and
LIRA, Observatoire de Paris, Université PSL, Sorbonne Université, Université Paris Cité, CY Cergy Paris Université, CNRS, 92190 Meudon, France
\and
Leiden Observatory, Leiden University, Einsteinweg 55, 2333, CC Leiden, The Netherlands
\and
Mullard Space Science Laboratory, University College London, Holmbury St Mary, Dorking, Surrey, RH5 6NT, UK
\and
Sydney Institute for Astronomy, School of Physics A28, University of Sydney, Australia
\and
Royal Observatory of Belgium, Ringlaan 3, 1180, Brussels, Belgium
\and 
Institute of Astronomy, University of Cambridge, Madingley Road, Cambridge CB3 0HA, UK
}

\abstract 
{The upcoming GaiaNIR mission is currently planning to add a near-infrared spectrograph to its payload in order to enhance its scientific return, particularly for mapping the dust-obscured regions of the Milky Way.} 
{We identify the optimal wavelength region between 800 and 2300 nm for the proposed GaiaNIR spectrograph to maximize the precision for radial velocities and atmospheric parameters.} 
{To determine its spectral range, we generated 10\,000 synthetic spectra from the BOSZ library across a wide range of stellar parameters, with resolutions varying from 5000 to 20\,000. By cross-correlating these mock observations with ideal templates, we assessed the statistical scatter of the velocity residuals to isolate six candidate windows for further testing of the atmospheric parameters.} 
{Our analysis found that the 1926 $-$ 1968 nm window at R = 16100 $-$ 20100 in the K band is the preferred strategic choice because it has the potential to reach a radial velocity precision for the brightest FGKM stars of about 160 $-$ 260 \ms \ depending on the resolution while providing a precision of the atmospheric parameters close to that of  the Gaia Radial Velocity Spectrometer. We also identified a second region between 1158 and 1202 nm (R = 9300 $-$ 11600) that has a slightly lower radial velocity precision, but at a wider temperature range than the K band. Both regions enable the derivation of abundances of ten species at these resolutions: O, Na, Mg, Si, Ca, Ti, V, Cr, Mn, and Ni in the K band and Mg, Si, K, Ca, Ti, V, Cr, Mn, Co, and Ni between 1158 $-$ 1202 nm.} 
{The K band delivers sufficiently precise measurements for the primary cool star targets of the mission while taking advantage of significantly lower interstellar extinction to enable the mapping of the dust-obscured regions of the Milky Way.}

\keywords{techniques: spectroscopic -- 
  Galaxy: abundances -- 
  Galaxy: evolution -- 
  Galaxy: fundamental parameters}

\titlerunning{GaiaNIR spectrograph}
\authorrunning{M{\'e}sz{\'a}ros et al. 2026}
\maketitle

\section{Introduction}

The launch of the Gaia mission \citet{2016A&A...595A...1G} in 2013 marked the beginning of a new era in all-sky space astrometry, photometry, and spectroscopy. Continuously operating from 2014 to 2025, the mission has successfully observed over two billion objects. The unprecedented volume of the derived positions, proper motions, parallaxes, radial velocities, atmospheric parameters (\teff, \logg, and \mh) and abundances of multiple elements has fundamentally transformed our understanding of the Universe. To date, Gaia has delivered three highly successful data releases, the first in 2016 \citep{2016A&A...595A...2G}, the second in 2018 \citep{2018A&A...616A...1G}, and the third in two parts in 2021 and 2023 \citep{2021A&A...649A...1G, 2023A&A...674A...1G}, with the fourth and fifth data releases scheduled for 2026 and 2030, respectively. This monumental success came from primarily mapping only about 1\% of the Galaxy that is heavily biased toward the solar neighborhood. Thus, the observed objects are severely constrained in space as Gaia operated in the optical wavelengths, which limited its ability to make observations of the inner regions Galaxy, the bulge, bar, and center regions, not to mention the other side of the Milky Way disk.  

Because much of our Galaxy is obscured by interstellar dust that seriously restricts our ability to map the Milky Way, a greater overall science return can only be achieved with a Gaia-like survey mission that expands the wavelength range to the near-infrared (NIR), where interstellar reddening is significantly lower than in the optical. This would allow us to obtain much deeper observations of the Galaxy than was ever possible before. Thus, a new all-sky NIR astrometric mission called GaiaNIR was proposed by \citet{2016arXiv160907325H, 2021ExA....51..783H}. GaiaNIR considers using MCT Avalanche Photo Diodes (APDs) that would allow the mission to perform astrometry and photometry between 800 and 2300 nm to expand and improve all science cases of Gaia. The short-wavelength limit is constrained by the detector sensitivity, while the long-wavelength cutoff avoids the necessity for active spacecraft cooling. 

As Gaia has already demonstrated, the science return of the mission can be greatly improved by the addition of a moderately high-resolution spectrograph. The Radial Velocity Spectrometer \citep[RVS, ][]{2018A&A...616A...5C} on Gaia covered a 27 nm wide window around the Ca II triplet lines (847$-$874 nm) at a moderately high average resolution of 11\,500. The most important discoveries of Gaia were greatly enhanced by the kinematical and chemical information derived from RVS spectra. The scientific impact of GaiaNIR might be exponentially enhanced by the inclusion of a high-resolution spectrograph (preferably R $>$ 10\,000) in the mission profile that covers a moderate-wavelength region to measure radial velocities, atmospheric parameters, and abundances for potentially hundreds of millions of stars. Because it is currently not technically feasible to build an R $>$ 10\,000 spectrograph for a space mission that covers the entire 800 $-$ 2300 nm region, the wavelength range must be limited to maintain the required high resolution of the spectrograph. 

The selection of the optimal window is a complex optimization problem that requires balancing high radial velocity precision with the ability to accurately extract atmospheric parameters and chemical abundances from stellar spectra. The latter requires higher signal-to-noise ratios (S/N), especially if the spectra are used for detailed abundance studies. The only near-IR high-resolution spectroscopic surveys so far, the Apache Point Observatory Galactic Evolution Experiment \citep[APOGEE, ][]{2017AJ....154...94M} and Milky Way Mapper \citep[MWM, ][]{2026AJ....171...52K} surveys observed nearly a million stars between 2011 and 2022, but the covered wavelength was limited to the H band between 1.51 and 1.69 \mum \ with a resolution of 22\,500. Their latest atmospheric parameters and radial velocities were published as part of the 19th data release of SDSS \citep{2022ApJS..259...35A, 2025AJ....170...96M}).

Because a comprehensive library of high-resolution near-infrared observed spectra does not yet exist, we used theoretical spectral grids with resolutions between R = 5000 and 20\,000 with a baseline spectral window of 100 nm. Our goal was to identify wavelength regions that are optimal for deriving the radial velocity, atmospheric parameters and chemical information of stars observed by GaiaNIR. The paper is structured as follows: Sec.~\ref{sec2} describes our motivation and method, Sec.~\ref{sec3} discusses our rationale of selecting the wavelength region for the GaiaNIR spectrograph, and Sec.~\ref{sec4} contains the conclusions.

\section{Motivation and method}
\label{sec2}

\subsection{Initial assumptions}

Our assumptions for the GaiaNIR spectrograph were closely modeled after the successful design architecture of the RVS \citep{2018A&A...616A...5C}. The RVS was configured with a sampling of \~3 pixels per resolution element and a central wavelength of 860.5 nm (847$-$874 nm), with a window length of roughly 1000 pixels on the detector. To maintain the spectral resolving power, the orientation of the dispersion direction was set along the scanning direction. We assumed that GaiaNIR will adopt a similar operational framework. For our theoretical calculations, we assumed that the maximum resolution that can be achieved depends on the selected central wavelength ($\lambda_C$), the wavelength range centered on $\lambda_C$ ($\lambda_{r}$), the total number of pixels available on the detector for this region ($N_R$), and the sampling per resolution element ($N_P$). In this case, this maximum resolution can be calculated with the following equation:

\begin{equation}
R = \frac{\lambda_C \cdot N_R}{\lambda_{r} \cdot N_P}.
\label{eq1}
\end{equation}

Alternatively, $\lambda_{r}$ for a fixed resolution and $\lambda_C$ can be calculated by rearranging the equation. For all of our tests, we fixed the number of pixels per resolution element to three, following the RVS design. It is important to note that our goal here is not to give a recommendation on the final hardware configuration of the spectrograph, but rather to evaluate the required resolution and window size for extracting precise radial velocities and atmospheric parameters.

\begin{figure}                          
\centering
\includegraphics[width=3.2in,angle=0]{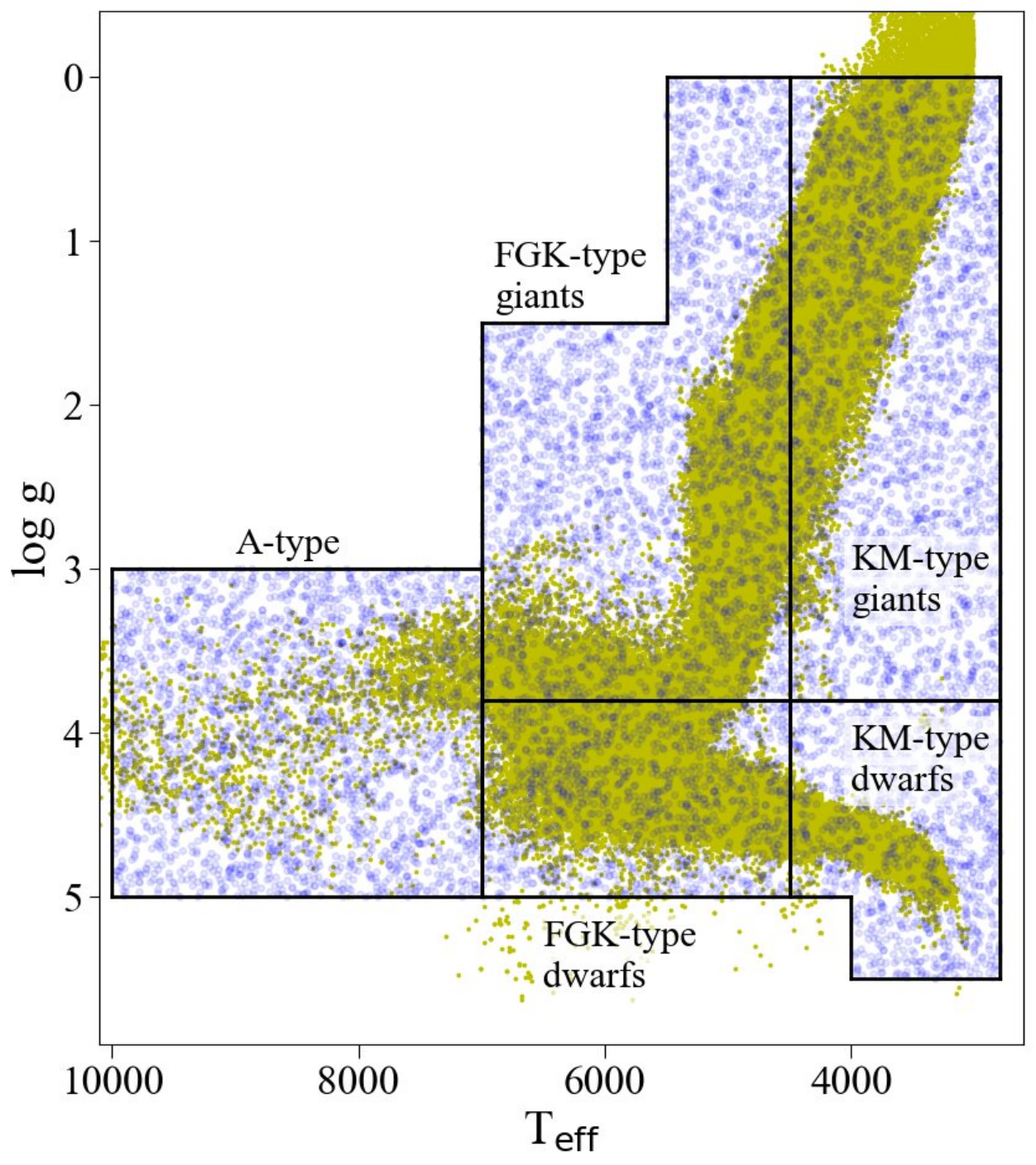}
\caption{Kiel diagram of the random models (blue dots) we used for the testing. We selected five regions to calculate the radial velocity and atmospheric parameter statistics, as indicated in the plot. Stars observed by MWM and APOGEE from DR19 are shown in the background with yellow dots for reference.}
\label{tefflogg}
\end{figure}

Our method was to first prioritize the wavelength regions that provide accurate and precise radial velocities. Consequently, we sequentially filtered our candidate windows: first, we isolated regions that provided robust radial velocities (by selecting the strongest absorption lines in the region to ensure a good \vrad \ precision), and we subsequently refined them based on the availability of spectral features suitable for atmospheric parameter and abundance determinations.

\begin{figure*}                          
\centering
\includegraphics[width=6.6in,angle=0]{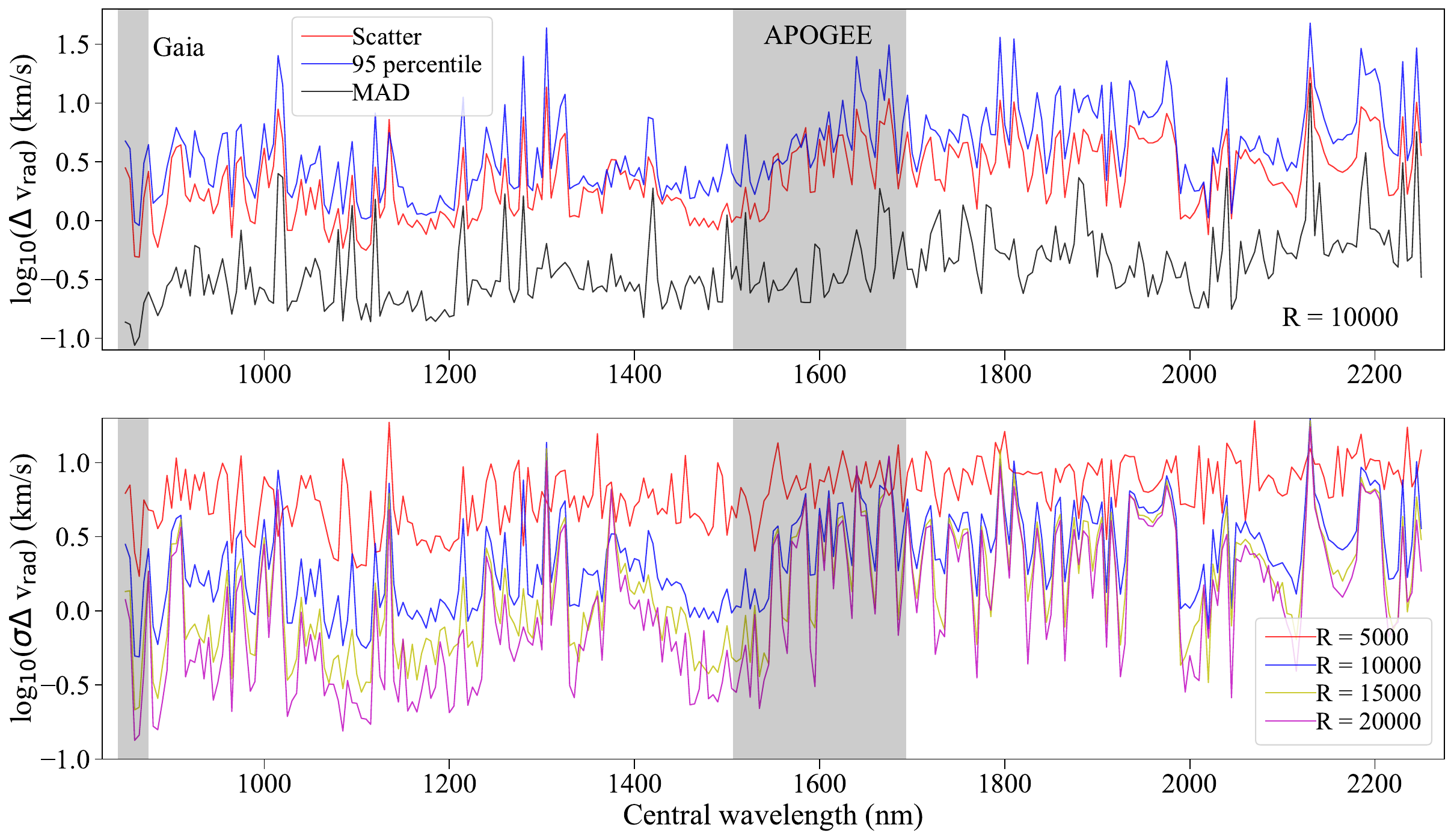}
\caption{Estimated precision of \vrad \ as a function of central wavelength using a wavelength region of 100 nm. The top panel shows the scatter, P$_{95}$, and median absolute deviation of the differences between the derived and assigned velocities for R = 10\,000. The bottom panel illustrates the variation in the scatter when the resolution changes from 5000 to 20\,000. The wavelength range covered by the Gaia and APOGEE spectrographs are highlighted by gray areas.}
\label{vrad}
\end{figure*}

We first focused on radial velocities because these values can reliably be derived at relatively low S/N values (even down to 10$-$15), which ensures that kinematic data are available for the vast majority of observed targets. Atmospheric parameters require a higher S/N, at least 30$-$40 to be reliable, while abundances demand even higher S/N values (at least 50$-$70 depending on the species). We finally compared the selected wavelength regions and suggest various resolutions for the design based on a number of factors: reddening, precision of the radial velocities, atmospheric parameters, and the species that are available in the spectrum. 

\subsection{Extinction}

Interstellar dust and extinction heavily obscure significant portions of the Milky Way. This severely restricted the ability of the original Gaia mission to map the inner galaxy and the opposite side of the Galactic disk. To overcome this limitation, the GaiaNIR mission is designed to operate in the near-infrared, where interstellar extinction and reddening are substantially lower. GaiaNIR is expected to observe six times more stars ($>$ 12 billion) than Gaia \citep{2021ExA....51..783H}, and this massive increase will primarily be concentrated in the Galactic disk, bulge, and star-forming regions in which dust is the thickest. If a suitable region at longer wavelengths where the extinction is as low as possible might be found for the GaiaNIR spectrograph, it would allow us to peer into these hidden dynamically important regions. GaiaNIR might finally capture the kinematic signatures needed to understand complex stellar movements (e.g., radial migration), the nature of dark matter halos, and the origin of the Milky Way spiral arms. Thus, we strongly prioritized longer wavelengths when regions with a similar radial velocity and atmospheric parameter precision were found.

\subsection{Simulating radial velocity measurements}

Because a comprehensive empirical library of high-resolution near-infrared spectra covering the entire 800$–$2300 nm wavelength range is currently unavailable, we used the updated BOSZ synthetic spectral database \citep{2024A&A...688A.197M}. These local thermodynamic equilibrium (LTE) models incorporate MARCS \citep{2008A&A...486..951G}  and ATLAS9 \citep{1979ApJS...40....1K, 2012AJ....144..120M} model atmospheres, updated continuous opacities, and 23 molecular line lists. The new grid was calculated with Synspec \citep{2021arXiv210402829H} using the LTE approximation and covers metallicities (\mh) from $-$2.5 to 0.75~dex, \am \ from $-$0.25 to 0.5~dex, and \cm \ from $-$0.75 to 0.5~dex, providing spectra for 336 unique compositions. The solar reference abundances are from \citet{Grevesse2007}.

To construct a suitable dataset for our simulations, we restricted the BOSZ grid to cover \teff \ between 2800 and 10\,000~K, \logg \ between 0 and 5.5~dex (depending on the temperature), and metallicity between $-$1.5 and 0.5~dex. The microturbulent velocity was fixed at 2 \kms, and four distinct resolutions were used: 5000, 10\,000, 15\,000, and 20\,000. These resolutions were kept the same between 800 and 2300 nm regardless of the wavelength. For the final tests, we randomly generated 10\,000 \teff, \logg, and \mh \ parameters in this restricted grid and interpolated the radial basis function to generate the corresponding theoretical spectra. The distribution of the random \teff-\logg \ pairs is illustrated in Fig.~\ref{tefflogg}.

To simulate realistic observations, random Gaussian noise was injected to achieve an arbitrary S/N value between 5 and 100. Each mock spectrum was subsequently Doppler-shifted by a random radial velocity between $-$100 and 100 \kms \ and cross-correlated with its original noise-free spectrum with the \texttt{crosscorrRV} function of \texttt{PyAstronomy\footnote{https://pyastronomy.readthedocs.io/en/latest/index.html}} from $-$150 to 150 \kms. This provided the ideal template for each spectrum because it excluded systematic errors from a template mismatch. 

The cross-correlation was performed across a 100 nm window, which we advanced in 5 nm increments from 850 nm to 2250 nm, yielding 281 discrete test regions. To evaluate the performance, the scatter ($\sigma$), median absolute deviation (MAD) and 95\% percentile ($P_{\rm 95}$) of the difference between the random \vrad \ and the new derived were calculated. These metrics were calculated for stars whose difference did not exceed $\pm$ 50 \kms \ to eliminate erroneous measurements, and these statistics were then used to assess the precision as a function of wavelength and resolution to determine the optimal wavelength region.

\subsection{Deriving the atmospheric parameters}

\begin{table*}[!t]
\begin{center}
\caption{Estimated precision of the radial velocity and atmospheric parameters.}
\begin{tabular}{lccccccccccc}
\hline
\hline
\multicolumn{1}{c}{$\lambda_C$} & 
\multicolumn{1}{c}{$\lambda_{min}$} & 
\multicolumn{1}{c}{$\lambda_{max}$} & 
\multicolumn{1}{c}{$\lambda_r$} & 
\multicolumn{1}{c}{$A_{\lambda} / A_{V}$} & 
\multicolumn{1}{c}{Window size} & \multicolumn{1}{c}{R} &
\multicolumn{1}{c}{$\sigma_{\rm v}$} & 
\multicolumn{1}{c}{$\sigma_{\rm Teff}$} & 
\multicolumn{1}{c}{$\sigma_{\rm log g}$} & 
\multicolumn{1}{c}{$\sigma_{\rm [M/H]}$} & 
\multicolumn{1}{c}{Available species} \\
\multicolumn{1}{c}{nm} & \multicolumn{1}{c}{nm} & \multicolumn{1}{c}{nm} & \multicolumn{1}{c}{nm} & 
\multicolumn{1}{c}{} & \multicolumn{1}{c}{pixels} & \multicolumn{1}{c}{} & 
\multicolumn{1}{c}{\kms} & 
\multicolumn{1}{c}{K} & 
\multicolumn{1}{c}{dex} & 
\multicolumn{1}{c}{dex} & 
\multicolumn{1}{c}{at high resolution} \\
\hline

1153 & 1134 & 1172 & 38 & 0.321 &  870 & 8800  & 0.563 & 155.1 & 0.644 & 0.089 & \textit{FGK}: Na, Mg, Si, K, Cr, Co, Ni \\
     &     &     &      &  & 1040 & 10500 & 0.410 & 144.5 &  0.59 & 0.086 & \textit{KM-giants}: Na, Mg, Al, Si, Ti, \\
     &     &     &      &  & 1300 & 13100 & 0.294 & 133.5 & 0.553 & 0.078 &  Cr, Co, Mn, Ni, CN \\
     &     &     &      &  & 1625 & 16400 & 0.197 & 127.8 & 0.531 & 0.076 & \textit{M-dwarfs}: Cr, Na, K \\   
     &     &     &      &  & 1950 & 19700 & 0.162 & 116.6 & 0.496 & 0.071 & \\
\hline
1180 & 1158 & 1202 & 44 & 0.31 &  870 & 7800  & 0.457 & 139.6 & 0.511 & 0.094 & \textit{FGK}: Mg, Si, K, Ca, Cr, Co \\
     &     &     &      &  & 1040 & 9300  & 0.335 & 128.4 & 0.458 & 0.079 & \textit{KM-giants}: Mg, Si, K, Ca, Ti, V, \\
     &     &     &      &  & 1300 & 11600 & 0.207 & 122.3 & 0.444 &  0.08 & Cr, Mn, Co, Ni, CN \\
     &     &     &      &  & 1625 & 14500 & 0.148 & 107.4 & 0.392 & 0.067 & \textit{M-dwarfs}: Mg, Si, K, Ca, Ti, Cr  \\
     &     &     &      &  & 1950 & 17400 & 0.123 &  99.4 & 0.366 & 0.063 & \\
\hline
1492 & 1474 & 1510 & 36 & 0.212 &  870 & 12000 & 0.847 & 254.0 & 0.375 & 0.122 & \textit{FGK}: Na, Mg, Si, Ti, Mn, Ni \\
     &     &     &      &  & 1040 & 14400 & 0.655 & 247.2 & 0.382 & 0.119 & \textit{KM-giants}: Na, Mg, Si, Ca, Ti, Cr \\
     &     &     &      &  & 1300 & 18000 & 0.398 & 218.9 & 0.357 & 0.105 &  Mn, Co, Ni, CN, OH \\
     &     &     &      &  & 1625 & 22400 & 0.339 & 194.0 & 0.363 & 0.098 & \textit{M-dwarfs}: Na, Mg, Ca, Ti,  \\  
     &     &     &      &  & 1950 & 26900 & 0.178 & 173.7 & 0.302 & 0.085 & Ba, Cr, Co, Mn, Ni \\
\hline
1947 & 1926 & 1968 & 42 & 0.138 &  870 & 13400 & 0.347 & 100.7 &  0.28 & 0.069 & \textit{FGK}: Na, Mg, Si, Ca \\
     &     &     &      &  & 1040 & 16100 & 0.261 &  77.3 & 0.201 & 0.061 & \textit{KM-giants}: Na, Mg, Al, Si, Ca, Ti, \\
     &     &     &      &  & 1300 & 20100 & 0.159 &  78.9 & 0.197 & 0.056 &   V, Cr, Ni, CN, OH \\
     &     &     &      &  & 1625 & 25100 & 0.136 &  94.9 &  0.27 &  0.06 & \textit{M-dwarfs}: Ca \\
     &     &     &      &  & 1950 & 30100 & 0.086 &  74.9 &  0.25 & 0.053 &  \\
\hline
1963 & 1926 & 2000 & 74 & 0.136 &  870 & 7700  & 0.681 & 116.8 & 0.394 & 0.081 & \textit{FGK}: Na, Mg, Si, Ca \\
     &     &     &      &  & 1040 & 9200  & 0.492 &  98.9 & 0.316 & 0.069 & \textit{KM-giants}: Na, Mg, Si, Ca, Ti, \\
     &     &     &      &  & 1300 & 11500 & 0.356 &  88.5 & 0.268 & 0.061 & V, Cr, Mn, Ni, CN, OH \\
     &     &     &      &  & 1625 & 14400 & 0.195 &  88.5 & 0.285 & 0.056 & \textit{M-dwarfs}: Ca \\
     &     &     &      &  & 1950 & 17200 & 0.139 &  84.0 & 0.263 & 0.052 & \\
\hline
1984 & 1968 & 2000 & 32 & 0.134 &  870 & 18000 & 0.148 & 169.7 & 0.683 & 0.089 & \textit{FGK}: Mg, Ca, Si \\
     &     &     &      &  & 1040 & 21500 & 0.118 & 157.5 &  0.63 & 0.083 & \textit{KM-giants}: Na, Mg, Si, Ca, Ti \\
     &     &     &      &  & 1300 & 26900 & 0.076 & 143.4 & 0.599 & 0.078 &  V, Cr, Mn, CN, OH \\
     &     &     &      &  & 1625 & 33600 & 0.061 & 134.1 & 0.566 & 0.077 & \textit{M-dwarfs}: Ca \\
     &     &     &      &  & 1950 & 40300 & 0.049 & 119.7 & 0.522 & 0.066 &  \\ 
\hline
\end{tabular}
\end{center}
\tablefoot{Estimated precision of the six selected windows assuming different window sizes in the detector, resulting in different resolutions. The estimated precision of \vrad \ is the scatter of the differences for stars with \teff \ $<$ 7000~K and S/N $>$ 40, and the precision of the atmospheric parameters was calculated using a more restricted sample of 4000 $<$ \teff \ $<$ 7000~K.}
\label{stats}
\end{table*}

To derive atmospheric parameters \teff, \logg, and \mh, we used the code FERRE \citep{2006ApJ...636..804A}, which interpolates a precomputed grid of synthetic spectra to find the stellar parameters that fit an observed spectrum best using a $\chi^2$ minimization algorithm. FERRE was run by fitting only the three parameters listed above, and the same statistics were then calculated as for the radial velocities. In order to keep the computation time feasible, we limited these atmospheric tests to the wavelength windows that were selected based on our \vrad \ tests in the previous step. For consistency, the same random spectra with the same noise were used in these tests as in the radial velocity step.

Four distinct theoretical grids were created in varying temperature and surface gravity regimes to facilitate an accurate FERRE fitting. The ranges of these grids were limited by the limits and step sizes of the APOGEE-MARCS model atmosphere database. The first grid covered temperatures from 2800 to 4000~K with steps of 100~K, \logg \ from 0 to 5.5 with steps of 0.5~dex. The second grid spanned a temperature from 4000 to 6000~K, surface gravities from 0 to 5, the third grid covered 5000 to 7500~K, and the fourth grid covered 6500 to 10500~K with \logg \ varying from 1.5 to 5 and 3 to 5, respectively. These three grids used 250~K \teff \ and 0.5~dex \logg \ steps. The microturbulence velocity was fixed to 2 km/s, and the same four resolutions were used as before. 

\subsection{Sample restrictions}

Because spectral features are highly sensitive to effective temperature, we divided the full sample into five subsamples based on temperature and surface gravity. They are shown in Fig.~\ref{tefflogg}, and they are slightly different from the grids we used to derive the atmospheric parameters to follow the standard spectral types more closely. The first of our subsamples covered red dwarf stars with \logg \ larger than 3.8~dex and \teff \ $<$ 4500~K. Red giants were defined as having \logg \ $<=$ 3.8~dex, and \teff \ $<$ 4500~K. This temperature limit was chosen because molecular lines usually start to dominate the spectra of stars below this temperature, depending on the wavelength. This group represents cooler K- and most M-type giants observed in the Milky Way. These cuts were chosen to reflect the specific spectral dominance of molecular bands versus atomic lines, which can significantly alter the precision of the radial velocity because molecular bands are only poorly resolved at low resolution.

To model F, G-, and warmer K-type stars, the temperature region between 4500 and 7000~K was also separated into two groups using the same surface gravity limit. The spectra of these stars are dominated by atomic absorption lines. The final subsample was defined as all stars with temperatures greater than 7000~K regardless of their \logg \ values. This was motivated by the fact that hot A-type stars are dominated by hydrogen lines and have very few metal lines, and thus, a wavelength region that is optimal for FGKM stars might not be optimal for hotter stars. During the evaluation process, we mainly selected our preferred wavelength windows based on the statistics measured using the FGKM samples because the vast majority of GaiaNIR targets will be cooler than 7000~K.

\begin{figure*}                          
\centering
\includegraphics[width=6.2in,angle=0]{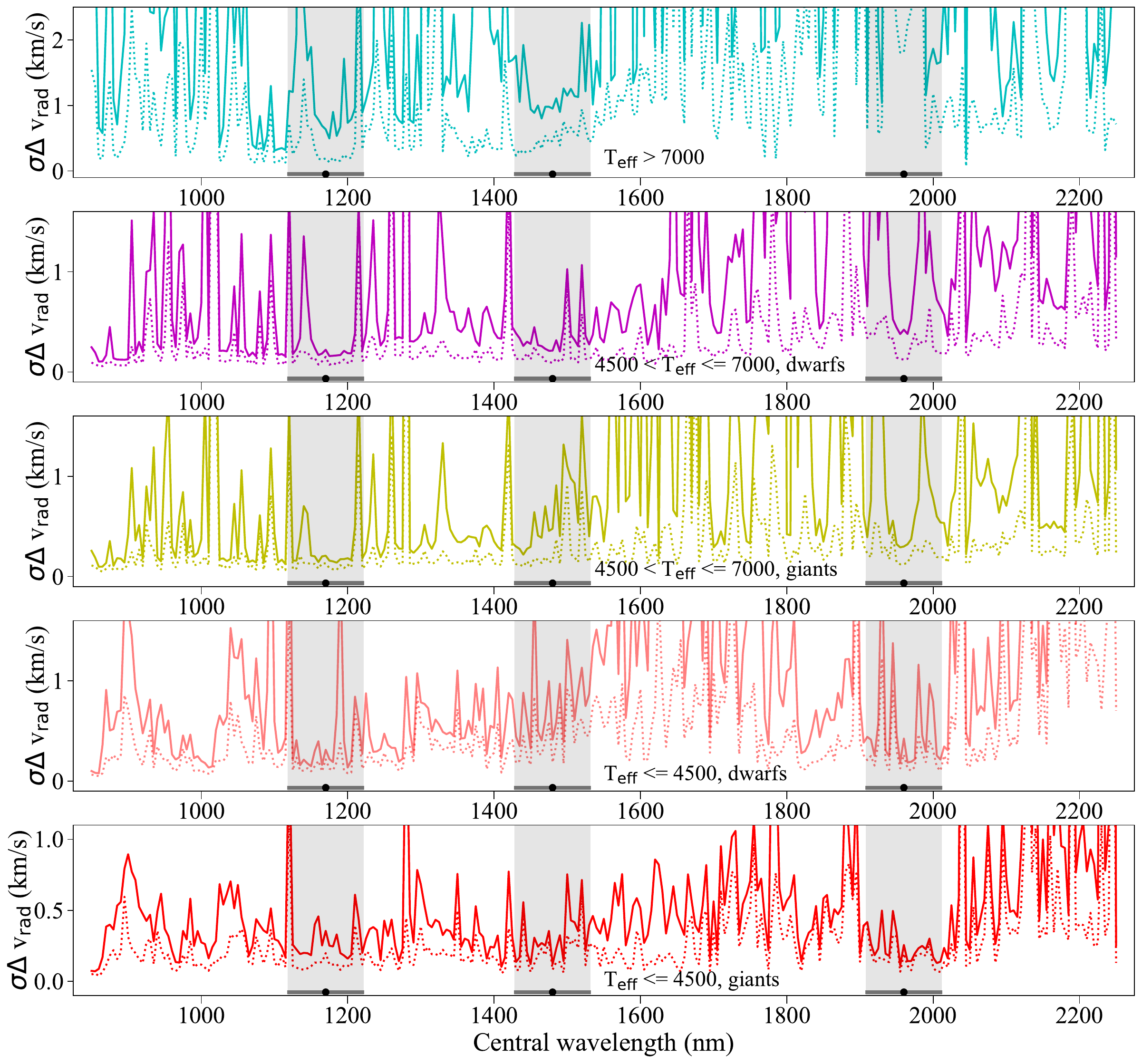}
\caption{Estimated average scatter in the radial velocity differences as a function of central wavelength. The gray areas highlight the initial three regions that were selected for the further analysis. These regions were selected to provide a low scatter of \vrad \ across a wide range of effective temperatures. The solid line shows the scatter, and the dotted line denotes the MAD.}
\label{radvelcomp1}
\end{figure*}

\subsection{Down-selecting windows}

The scatter of \vrad \ differences as a function of the central wavelength for the full sample is shown in Fig.~\ref{vrad}. It is important to note that the numbers shown in that figure and listed in this section do not indicate the final precision that can be achieved with GaiaNIR, but act as a guide to assess how these wavelength windows compare relative to each other, as all calculations were done consistently between windows. 

\begin{figure*}                          
\centering
\includegraphics[width=6.1in,angle=0]{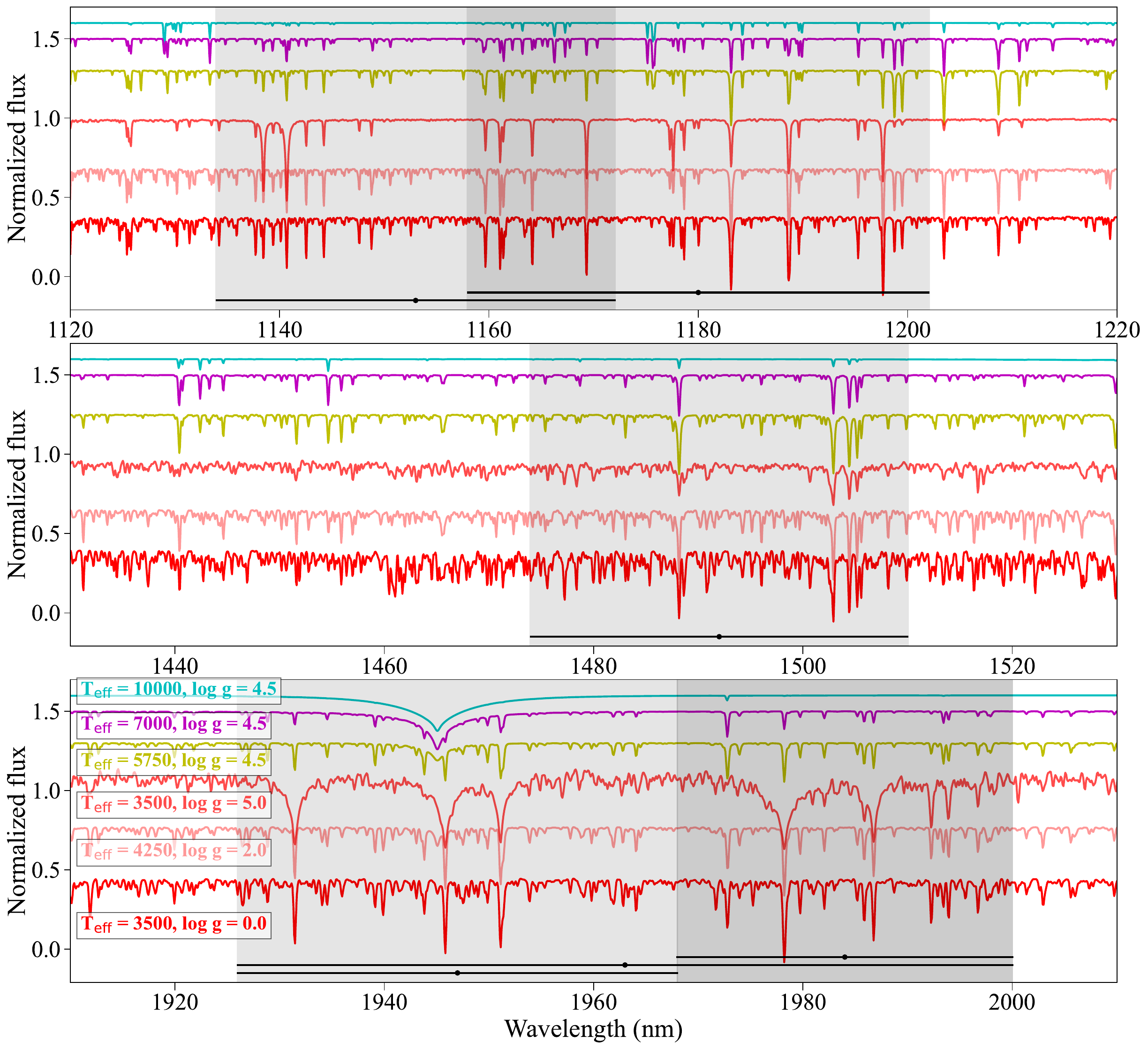}
\caption{Normalized flux of sample spectra in the three 100 nm windows selected in the first phase. Six windows were chosen for the further analysis of atmospheric parameter precision, these are shaded gray in the figure. The normalized flux was shifted by arbitrary numbers to aid visibility.}
\label{sp1}
\end{figure*}

\begin{figure*}                          
\centering
\includegraphics[width=6.3in,angle=0]{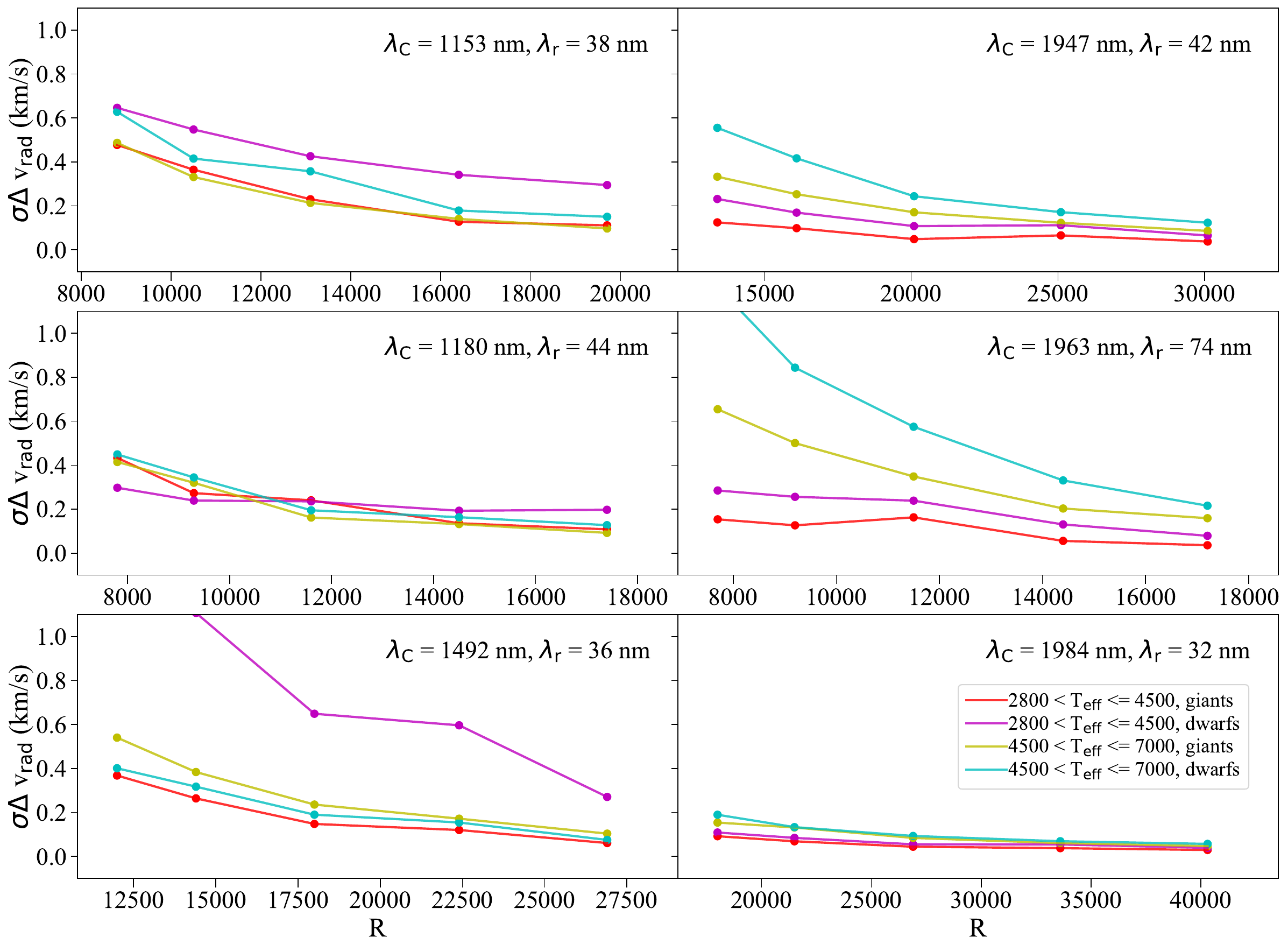}
\caption{Average scatter in the radial velocity differences as a function of R for the six wavelength windows selected in Section 2. The S/N $>$ 40 in all panels.}
\label{radvelcomp2}
\end{figure*}

Fig.~\ref{vrad} clearly demonstrates that multiple 100 nm wide windows are promising for measuring precise radial velocities between 800 and 2300 nm because in several regions, the scatter, MAD, and $P_{\rm 95}$ are low, notably for central wavelengths around 850$-$900, 960$-$980, and 1060$-$1200 nm, in the bluer part of the H band between 1450 and 1550 nm, and in the K band near 2000 nm. From these, we identified several regions for further refinement. Strong Ti lines between 960 and 980 nm offer precise radial velocities for K- and M-type stars, but become too weak for spectral types of AFG. A large group of Si, Ti, Fe, and other metallic lines between 1060 and 1090 nm provide good velocities for AFG-type stars, but not for stars below 4500~K. We therefore omitted these two regions from the further analysis.

The analysis of the five subsamples in Fig.~\ref{radvelcomp1} reveals a complex picture of the distribution of \vrad \ differences because the spectral lines are very sensitive to \teff. Of the full wavelength space, the 810$-$910 nm region ($\lambda_{\rm C}$ = 860 nm) provides the highest intrinsic precision in all subsamples, mirroring the success of the Gaia RVS. This enables us to determine precise radial velocities for all types of stars covering a wide temperature and metallicity range. However, selecting a window this close to the optical regime defeats the primary purpose of GaiaNIR because it fails to mitigate interstellar dust extinction. Consequently, we discarded this region.

Since GaiaNIR is expected to observe very many cool red giant and dwarf stars, it is more important to achieve a good precision for these two types of stars than for warmer OBA-type stars. Moreover, APDs are not sensitive to wavelengths shorter than 800 nm, which also limits the observation of hot stars. Overall, longer wavelengths are generally preferred over shorter ones to limit the effect of interstellar reddening and allow us to observe more stars. Taking all these effects into account, we selected three 100 nm wide windows for further consideration and focused on short-, medium-, and long-wavelength regions based on a careful assessment of how the scatter behaves in these various types of stars.

The first choice covers 1120$-$1220 nm, where the \vrad \ precision is very similar to that in the $\lambda_{\rm C}$ = 860 nm window for cool stars. This wavelength range performs worse than $\lambda_{\rm C}$ = 860 nm for stars with \teff \ $>$ 7000~K (top panel of Fig.~\ref{radvelcomp1}) because the absorption lines become weaker than near the Ca triplet. One disadvantage of this region is that the spectra contain no hydrogen lines and that almost all absorption lines disappear from the spectrum when the temperature reaches 12\,000$-$15\,000~K, making it impossible to measure radial velocities above these temperatures. This region benefits from a lack of severe molecular absorption, which simplifies continuum normalization, especially for the coolest stars. Unfortunately, the extinction near $\lambda_{\rm C}$ = 1170 nm (A$_{\rm V}$ / A$_{\rm \lambda}$ $\approx$ 3.22, \citet{1989ApJ...345..245C}) cannot be substantially lower than near $\lambda_{\rm C}$ = 860 nm, so selecting a window in this wavelength region would limit its strategic value for mapping obscured Galactic structures with GaiaNIR.

The bluer part of the H band between 1450 and 1550 nm again offers multiple good options. Fig.~\ref{vrad} shows that the radial velocities are more precisely measured below 1500 nm than where the APOGEE spectrograph operates. While APOGEE has already demonstrated that the H band offers exceptional velocities with an average precision of 50$-$70 \ms \ for RGB stars \citep{2025AJ....170...96M}, we did not choose this region because our simulations indicate that precise velocities in this regime require resolving powers exceeding R = 15\,000$-$20\,000 (see the bottom panel of Fig.~\ref{vrad}). This is because the molecular lines are too strongly blended at lower resolutions, and APOGEE has exceeded the resolution and the wavelength range available for GaiaNIR, which is limited by the detector size and stellar crowding. Although this region appears to be less precise than the previous two, a decent compromise between precision and extinction might be achieved by selecting a narrow window in the region from 1430 to 1530 nm ($\lambda_{\rm C}$ = 1480 nm).

At the longest wavelengths, strong Ca lines in the region between 1910 and 2010 nm ($\lambda_{\rm C}$ = 1960 nm) dominate the spectra and offer an excellent opportunity for radial velocity measurements. This region delivers $v_{rad}$ precision for cool giants and dwarfs that rivals the 1170 nm window at a fixed resolution, but the scatter increases more strongly for warmer stars (\teff \ $>$ 4500~K) than in the previous two regions. All these tests were performed at the same resolution, not at the same resolving power, which means that the longer wavelength allowed higher resolutions to a level that would allow a precision similar to or even better than the current Gaia precision for all FGKM stars. Crucially, the interstellar reddening in the K band is approximately 7.4 times lower than at optical wavelengths \citep{1989ApJ...345..245C}, and thus, operating in this regime would fundamentally fulfill the GaiaNIR mandate to map the thin disk and even the far side of our Galaxy.

\section{Radial velocities, atmospheric parameters, and abundances}
\label{sec3}

The initial 100 nm sliding window analysis served as an idealized search to find information-rich spectral regions. However, implementing 100 nm wide wavelength regions with resolutions close to or exceeding R = 10\,000 is unlikely to be feasible given the anticipated detector size limitations and telemetry constraints. Therefore, we further restricted the range of our three primary candidate windows and prioritized the retention of the strongest absorption lines to preserve the \vrad \ precision while discarding relatively featureless continuum regions.

We evaluated five potential configurations: a baseline Gaia-like configuration using 1040 pixels to sample the spectrum, one profile using fewer pixels (yielding a lower resolution), and three profiles using more pixels (yielding higher resolutions). This allowed us the potential to measure more precise parameters and abundances at the cost of an increased size on the detector. The covered spectral regions were fixed in all five cases for each central wavelength, as listed in Table~\ref{stats}. These resolutions were calculated using Equation~1. We also examined how precisely atmospheric parameters (\teff, \logg, and \mh) can be derived from the down-selected regions, and we explain in detail which species are available from them. The wavelength space of the final recommendation is shown in Fig.~\ref{sp1}. Table~\ref{stats} presents the realizable performance with the fixed-sized pixels we used to sample the observed spectra, our estimated precision of the radial velocity with an S/N $>$ 40 and 2800 $<$ \teff \ $<$ 7000~K, and the atmospheric parameters with the same S/N cut, but sampling stars with 4000 $<$ \teff \ $<$ 7000~K. The estimated average scatter of \vrad and the atmospheric parameters are shown in Fig.~\ref{radvelcomp2} as a function of resolution for the six down-selected windows with the five different configurations detailed in this section.

\subsection{Validation of our estimated precision}

To validate our theoretical method, we performed the same calculations in the simulated 845$-$872 nm range with a resolution of 11500 and compared our estimated uncertainties to the empirical performance of Gaia data release 3. \citet{2023A&A...674A..29R} estimated the precision of \teff, \logg, and \mh \ in DR3 by comparing Gaia parameters with those of APOGEE DR17, GALAH DR3, and RAVE DR6. Using the best-quality subsample, they found that the standard deviations of the differences were 90.3~K of \teff, 0.19~dex of \logg, and 0.13~dex for \mh. By selecting the same parameter range (4000 $<$ \teff \ $<$ 7000~K, 1 $<$ \logg \ $<$ 5, and $-$1 $<$ \mh \ $<$ 0.6) from the theoretical library we used, our estimated precisions are 83.1~K, 0.15~dex, and 0.06~dex for the same parameters. Considering that the numbers reported by \citet{2023A&A...674A..29R} also folded in errors from the comparison surveys, we can conclude that our estimated precisions closely replicate that of the observational scatter, although our metallicity estimates might be slightly optimistic.

It was harder to verify our estimated uncertainty of \vrad \  because \citet{2023A&A...674A...5K, 2023A&A...674A...6S} gave the calculated median formal precision, not the standard deviation of the differences between known and fitted values, as we do in this study. The most precise radial velocities in Gaia DR3 are achieved when \snr \ $>$ 200. For giant stars with \logg \ $<$ 3.5 and \teff \ $<$ 5000~K, Gaia DR3 has an estimated uncertainty of around 1.4$-2.0$ \kms \ at \snr \ $\approx$ 10, 360$-$440 \ms at \snr \ $\approx$ 40, and 175$-$200 \ms at \snr \ $\approx$ 100, which flattens out at 120$-$140 \ms \ for the brightest stars. Our estimated precision is 157 \ms \ for the same \logg \ range at an S/N $\approx$ 100, 300 \ms \ at \snr \ $\approx$ 40, and 680 \ms \ at S/N $\approx$ 10. This comparison suggests that when \snr \ $>$ 40, our estimated precision is slightly optimistic but close to the observed precisions, but at low \snr \ values, we underestimate it by about a factor of 2$-$3 at least.

For the main sequence stars (\logg \ $>$ 3.5), the Gaia DR3 precision strongly depends on temperature. Between 4000 and 7000~K, the DR3 uncertainty is around 2.2$-3.1$ \kms \ at \snr \ $\approx$ 10, 560$-$770 \ms at \snr \ $\approx$ 40, 220$-$300 \ms at \snr \ $\approx$ 100, which flattens out at 120$-$150 \ms \ for the brightest stars. The scatter we measure in the same temperature range is about 840 \ms \ at S/N $\approx$ 10, 230 \ms \ at \snr \ $\approx$ 40, and 160 \ms \ at \snr \ $\approx$ 100. Similarly to the giant stars, at high \snr \, our estimated uncertainty is slightly optimistic, but as \snr \ decreases, we differ more strongly from the observations. For MS stars below 4000~K, the precision of Gaia DR3 \vrad \ is about 350 \ms \ at \snr \ $\approx$ 100, 610 \ms \ at $\approx$ 40, and 2.7 \kms \ at $\approx$ 10, while our estimates are 440 \ms, 450 \ms, and 800 \ms, respectively.

Based on these values, we can conclude that the estimated precision  calculated here from the scatter of the radial velocity differences reproduces the observations fairly well for the brightest stars with the highest \snr and underestimates them for low S/N values. Thus, the results we presented for radial velocities in the wavelength ranges used here might give realistic uncertainties for the bright stars, which provide the best information that can be extracted, but the bulk of GaiaNIR targets will be fainter. While with this study, our goal was to find wavelength regions that are excellent for determining the radial velocity and atmospheric parameters, a more detailed study of the expected radial velocity uncertainties at low \snr \ remains to be explored in a later study.

\subsection{Abundance calculations}

We identified elements that produce measurable abundance signatures within the proposed wavelength coverage of the spectrograph. The following three synthetic spectra at solar metallicity were calculated to determine the variations in the elemental abundances that generate detectable spectral features in these regions: 1. a G dwarf: \teff \ = 5777~K and \logg \ = 4.438. 2. a K giant: \teff \ = 4000~K and \logg \ = 1.0. 3. an M dwarf: \teff \ = 3500~K and \logg \ = 5.0.

The synthetic spectra were computed using MARCS \citep{2008A&A...486..951G} model atmospheres in combination with the radiative transfer code \texttt{Turbospectrum}, assuming LTE for all elements. Atomic line data were adopted from \texttt{VALD3} \citep{Ryabchikova2015}, and the molecular line lists were compiled from a combination of \texttt{VALD3} and a line list kindly made available by B.~Plez. Microturbulent velocities of 1~km~s$^{-1}$ were adopted for the dwarf models and 2~km~s$^{-1}$ for the giant model, following standard MARCS prescriptions for these stellar types. All calculations were performed in vacuum wavelengths, adopting solar reference abundances from \citet{Grevesse2007}. The resulting spectra were continuum-normalized and convolved with a Gaussian kernel in logarithmic wavelength space to reproduce the proposed instrumental resolving powers in Table~\ref{stats}.

For each element, the abundance was perturbed by $+0.2$~dex while all other elemental abundances were held fixed at their solar values. This perturbation scale was chosen to represent a practical benchmark for assessing detectability. A spectral feature was considered detectable when this perturbation produced a flux difference exceeding the $1\%$ level relative to the solar abundance spectrum, corresponding to the detection threshold for a signal-to-noise ratio of S/N~$=100$.

\begin{figure*}                          
\centering
\includegraphics[width=7.0in,angle=0]{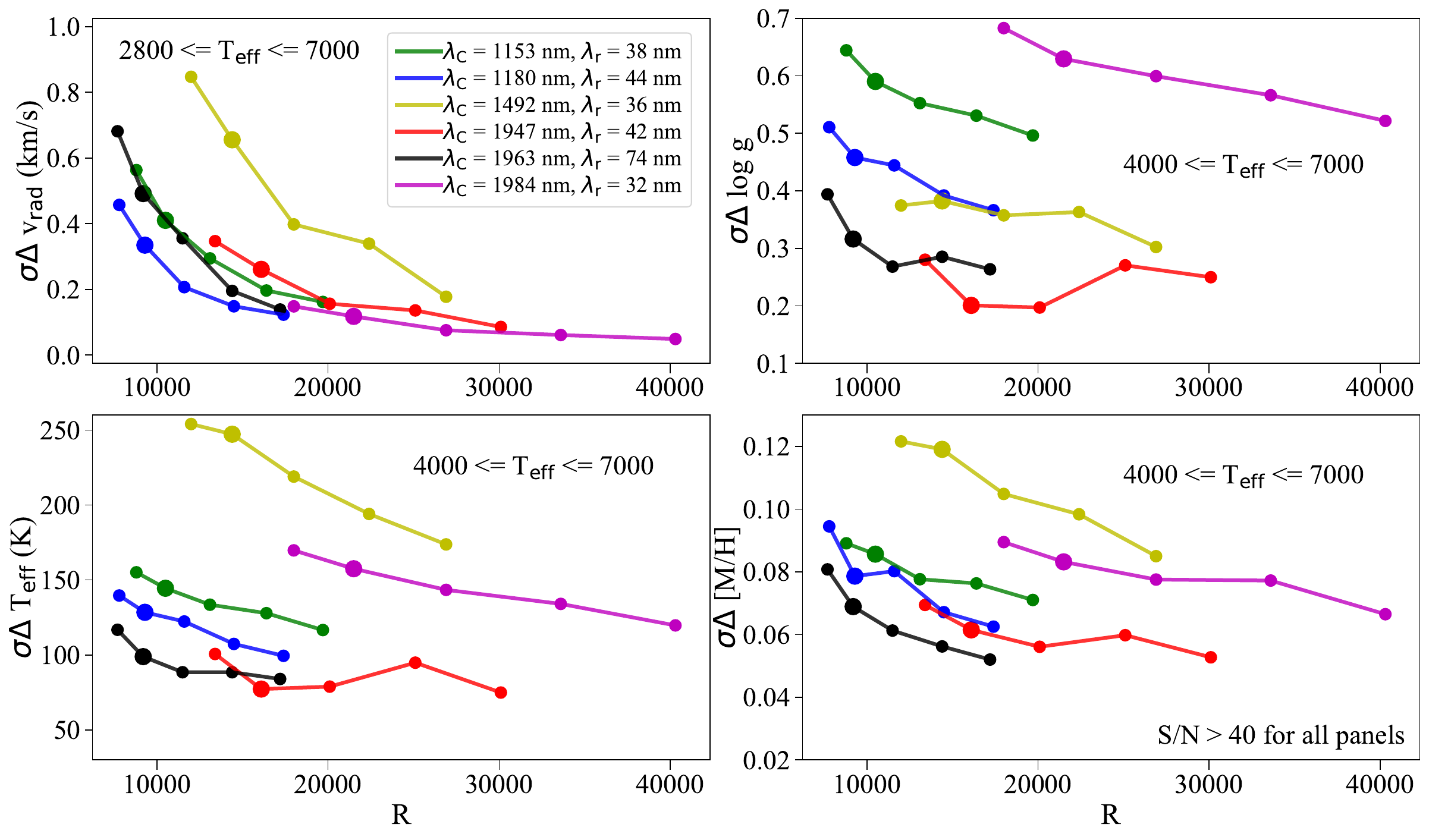}
\caption{Average scatter in the differences of \vrad, \teff, \logg, and \mh \ as a function of R for the six wavelength windows selected in Section 2. Each colored line represent a different down-selected window listed in Table~\ref{stats}. The large filled circles in each line denote a Gaia-like configuration using 1040 pixels to sample the spectrum. The precision of \vrad \ was estimated using a wider \teff \ range than the atmospheric parameters, as indicated in each panel.}
\label{apcomp}
\end{figure*}

\subsection{$\lambda_C$ = 1153 and 1180 nm}

We examined two restricted windows between 1120$-$1220 nm with a focus on the strongest absorption lines. Of these two windows, the window centered on 1180 nm consistently outperformed the 1153 nm window. It provided a better precision for radial velocity and all atmospheric parameters regardless of resolution (Fig.~\ref{apcomp}). This was reinforced by confirming the distribution of the differences in these two regions as a function of \teff \ in Figures~\ref{radveldiff}$-$\ref{apcomp3}. For FGKM stars, the average estimated scatter is 335 \ms \ in the $\lambda_C$ = 1180 nm range compared to 410 \ms \ for $\lambda_C$ = 1153 under a Gaia-like configuration (\snr \ $>$ 40). We examined the BOSZ theoretical spectra and found that absorption lines are still visible at temperatures up to about 12\,000$-$16\,000~K, making it possible to derive radial velocities for hot stars, although at a lower precision. 

Upon further examination of the $\lambda_C$ = 1180 nm range, the estimated precision of the effective temperatures might be somewhat higher than what was possible with Gaia. Assuming a Gaia-like configuration (which gives R = 9300 at $\lambda_C$ = 1180 nm), we estimate the scatter of \teff \ to be 128.4~K, while the same method gave 83.1~K for the RVS. The determination of the surface gravity seems to be even more unreliable, as our scatter is 0.458~dex compared to the 0.15~dex we calculated for Gaia. This strongly signals that the $\lambda_C$ = 1180 range might not provide precise \logg \ values because the spectral lines in this region are not sensitive to the surface gravity. The situation is much better for the metallicity, for which we estimate that the precision might be on par with Gaia, around 0.08~dex for FGKM stars.

\begin{figure*}                          
\centering
\includegraphics[width=6.5in,angle=0]{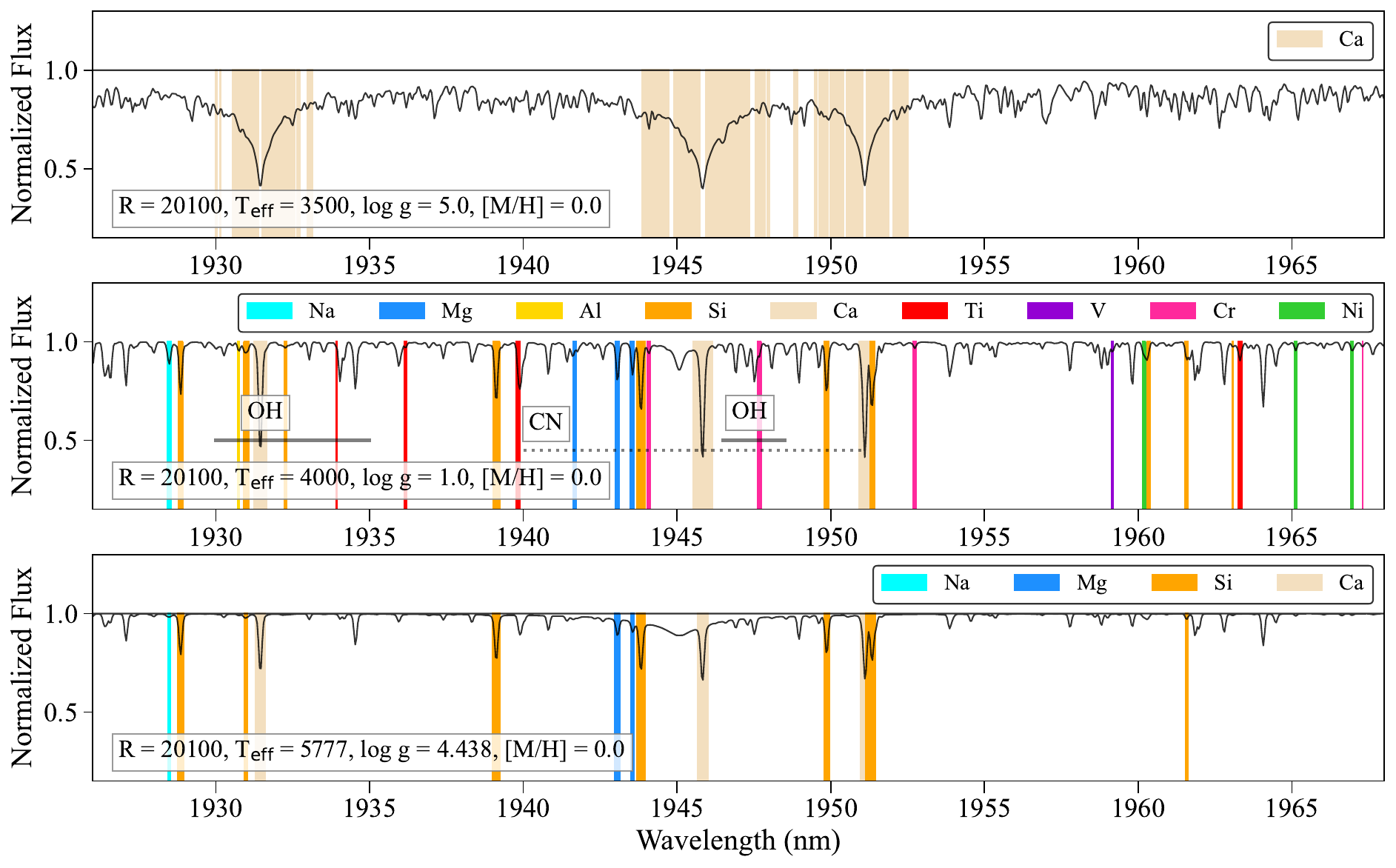}
\caption{Elements in the $\lambda_C$ = 1947 nm window with residuals larger than 0.01 for an abundance change of 0.2 dex in the selected synthetic spectra, as indicated in the figure.}
\label{elems2}
\end{figure*}

This $\lambda_C$ = 1180 region is highly advantageous for continuum normalization because it contains no significant molecular absorption lines, as only CN has moderately strong absorption lines in this region at higher temperatures, while water only becomes significant below 3000$-$3200~K. Our tests for a 3500~K, \logg \ = 5.0 red dwarf show that measuring abundances of Mg, Si, K, Ca, Ti, and Cr is possible for the majority of M dwarfs (see Table~\ref{stats}, and Figures~\ref{elems1}$-$\ref{elems2}). The spectra of FGKM stars are dominated by strong neutral Fe, Mg, Si, and Co lines with weaker lines of K, Ca, and Cr. In case of red giants, the list of available elements expands, with V, Mn, and Ni. Overall, this wavelength region offers a rich variety of ten species across a wide range in temperature and \logg. These elements allow rich science, for example, tracing the Milky Way accretion history, mapping star formation histories, or investigating globular cluster dissolution. N might be measured from the CN lines, on the other hand, but only if the amount of C is known. However, this wavelength region does not contain measurable CO lines, and thus, C can only be estimated using photometry.

Overall, the  $\lambda_C$ = 1180 range provides an excellent \vrad \ and metallicity precision. This is on par with Gaia (especially if more pixels can be used to increase the resolution above 10\,000), with a slightly decreased \teff \ precision. Unfortunately, measuring \logg \ appears to be severely limited in this region.  \vrad \ might be derived to relatively high temperatures of around 12\,000$-$16\,000~K, \teff \ up to about 8000$-$9000~K and the metallicity up to 9000~K (see Fig.~\ref{apcomp1}, ~\ref{radveldiff} and \ref{apcomp3}). However, this region covers only slightly longer wavelengths than the spectrograph on the Gaia mission, so it does not offer a substantial reduction in interstellar extinction. This might prevent GaiaNIR from fully taking advantage of the NIR wavelength and properly probing the dust-obscured areas of the Milky Way. We therefore preferred to search for more suitable windows at longer wavelengths.

\subsection{$\lambda_C$ = 1492 nm}

While our initial tests showed that precise radial velocities can be measured between 1430 and 1530 nm, tests on the more restricted window centered on the strongest lines at 1492 nm (1474 $-$ 1510 nm) revealed that this region exhibits the poorest precision among all six down-selected candidates. Our estimated \vrad \ scatter for FGKM stars is 655 \ms \ for a Gaia-like configuration (R = 14\,400, \snr \ $>$ 40), approximately a factor of three worse than current Gaia RVS capabilities. The precision significantly improves with an increased resolution and a larger window size on the detector. If sampling 1950 pixels is possible (increasing the resolution to 26900), the precision of \vrad \ improves to 178 \ms. While this is excellent, it may come at the price of significant crowding because the spectra on the detector are larger. This agrees with the tests we presented in Section~2.5: the wavelengths around the H band only provide precise velocities at a high resolution and in a wide wavelength range.

Unfortunately, the precision of the atmospheric parameters is similarly degraded. For a Gaia-like configuration, we estimate that \teff \ can be measured with an average precision of 247.2~K and a metallicity at 0.119~dex, both of which fall short of the Gaia performance. In case of \logg \,, there is some improvement over the $\lambda_C$ = 1180 nm range, but it still falls short of the Gaia precision. Most of the strong absorption lines belong to Fe, Na, Mg, Si, and Ca in FGKM stars, with Ti, Cr, Mn, Co, Ni, and Ba having weaker lines. The primary chemical advantage of this region is the barium line in the spectrum, which is the only available r-process element in all of our six test windows. Most molecular lines originate from OH and CN, but H$_{\rm 2}$O becomes a significant absorber below 4000~K. Atomic absorption lines disappear from the spectra above 9000$-$10\,000~K, making it very difficult to measure radial velocities or atmospheric parameters of OBA stars.

When we take the decreased ability to measure radial velocities and atmospheric parameters into account, the $\lambda_C$ = 1492 nm range is not recommended for observation. However, more wavelength windows near the H band can be profitably explored if APDs are found to have a cutoff below 1800 nm, to determine whether it is possible to find a suitable window that can provide at least Gaia-like precision.

\subsection{$\lambda_C$ = 1947, 1963, and 1984 nm}

As we found in Sec.~2.5, strong Ca lines in the region between 1910 and 2010 nm dominate the spectra, making it highly attractive for kinematic studies. Because these Ca lines cover a relatively wide wavelength range, we subdivided this region into three specific windows (see Table~\ref{stats}). The window centered on 1963 nm covers all the major Ca lines between 1926 and 2000 nm, offering a medium resolution, R = 9\,200, for a Gaia-like configuration. Narrower windows focusing on the shorter and longer wavelength range were also defined, making it possible to explore higher resolutions. The first window was centered on 1947 nm (1926 $-$ 1968 nm) and offered a resolution of 16\,100, while the second window covered the wavelength range of 1968 to 2000 nm ($\lambda_C$ = 1984 nm) with a resolution of 21\,500 in case of a Gaia-like setup.

For a Gaia-like configuration, the $\lambda_C$ = 1984 nm window provides the precisest velocities at 118 \ms \ of all the down-selected six windows. This might suggest a potential improvement over Gaia for the brightest FGKM stars and highest S/N observations. The $\lambda_C$ = 1947 nm range offers a somewhat lower precision at 261 \ms, which is still on par with Gaia. The precision decreases for the full range available around $\lambda_C$ = 1963 nm, where the estimated scatter increases to 492 \ms \ at R = 9\,200. This is clearly the effect of the decreased resolution because the absorption lines in the K band are often blended at resolutions around 10\,000. Fig.~\ref{apcomp} demonstrates this effect and clearly shows that the scatter drops sharply as the resolution increases. At R = 14\,400, the scatter drops to 195 \ms, which is now on par with what Gaia was capable of for the brightest FGKM stars.

\begin{figure*}                          
\centering
\includegraphics[width=6.5in,angle=0]{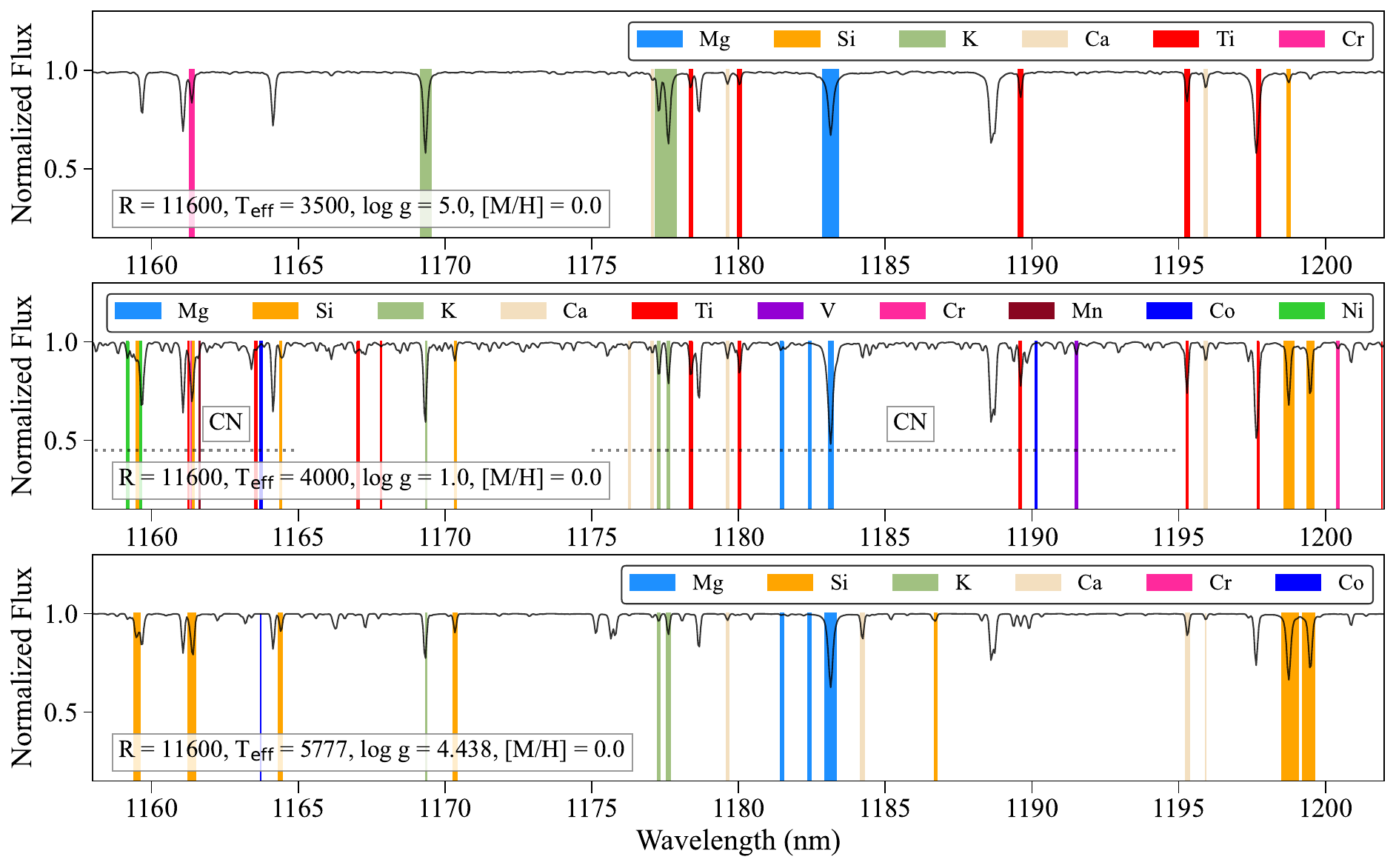}
\caption{Elements in the $\lambda_C$ = 1180 nm window with residuals larger than 0.01 for an abundance change of 0.2 dex in the selected synthetic spectra, as indicated in the figure.}
\label{elems1}
\end{figure*}

While the $\lambda_C$ = 1984 nm window performed best for the radial velocity, the opposite is true for \teff, \logg, and \mh, for which this region provides the poorest performance in the K band. The scatter of 157.5~K for the temperature is not only worse than what Gaia achieved, but also higher by about a factor of two than what the $\lambda_C$ = 1947 and 1963 nm windows can offer at 77.3~K and 98.9~K, respectively. These two values are very close to what we estimated for Gaia (83.1~K). For surface gravities, we estimate a precision of 0.201~dex and 0.316~dex for the $\lambda_C$ = 1947 and 1963 nm windows, respectively, which are somewhat higher values than was possible with Gaia. Fig.~\ref{apcomp} shows that the precisions of \logg \ and \teff \ do not seem to improve with increasing resolution for the $\lambda_C$ = 1947 and 1963 nm windows for window sizes larger 1300 pixels, signaling that surface gravity and effective temperature measurements can no longer be improved above R $>$ 11\,000$-$12\,000 for $\lambda_C$ = 1963 nm and R $>$ 20\,000 for $\lambda_C$ = 1947 nm. 

The decreased precision of the atmospheric parameters in the $\lambda_C$ = 1984 nm region can significantly affect the overall science return of the mission. On one hand, a lower \teff \ precision will lead to poorer fundamental stellar parameters, affecting stellar evolution studies that determine the mass of stars, and so on. On the other hand, the uncertainties in the atmospheric parameters will increase the uncertainty of the abundances through standard error propagation laws, which can prevent us from making detailed chemical abundance studies of the inner Milky Way and the far side of the Galactic disk. While the $\lambda_C$ = 1984 nm region gives the most precise \vrad \ values, a decent compromise might therefore be achieved by selecting the $\lambda_C$ = 1947 nm region with a resolution of $<$ 20\,000, because it provides atmospheric parameters that are more precise by nearly a factor of two at the expense of radial velocities that are less precise by about a factor of two, and even the precision of these radial velocities will be on par with that of Gaia.

For the metallicity, windows centered on 1947 and 1963 nm provide the precisest values, 0.061 and 0.069~dex, respectively. This is slightly better than the $\lambda_C$ = 1984 nm window. This performance is very similar to that of Gaia, as for that mission we estimated a precision of \mh \ 0.06~dex with the method used in this paper. Measuring metallicities will be extremely difficult in the K band in one temperature range,, and that is below about 3700$-$3800~K (see Fig.~\ref{apcomp3}). For these cool M dwarfs, the lines of H$_{\rm 2}$O overwhelm the spectrum, making the spectral lines insensitive to the metallicity and temperature (see Fig.~\ref{apcomp1}). This means that accurate atmospheric parameters for the coolest M dwarfs will be significantly challenging to determine from these windows.

Above 8000~K, the $\lambda_C$ = 1947 and 1963 nm windows both contain one hydrogen line (Br$-\delta$) and exhibit very few metallic lines. The radial velocities for OBA-type stars can therefore be measured with only a relatively poor precision (see the bottom left panel of Fig.~\ref{radveldiff}). These two regions would also allow us to estimate \teff \ from this single hydrogen line, an advantage lacking in the $\lambda_C$ = 1984 nm window. Below 7000~K strong lines of Fe, Mg, and Ca and moderately strong lines of Si are visible in the spectrum of solar-like stars with one weak Na line (Fig.~\ref{elems2}). At very low temperatures, the lines of H$_{\rm 2}$O completely dominate the spectrum, and only the strong Ca lines remain. Thus, below 4000~K, the abundance of Ca alone can be determined with current methods. The number of available species increases significantly for red giant stars, as strong absorption lines of Na, Mg, Al, Si, Ca, Ti, and Ni occur in the spectra (Fig.~\ref{elems1}), with V and Cr having only weaker lines. The $\lambda_C$ = 1947 and 1963 nm ranges cover Al, but not Mn, which only becomes available in the $\lambda_C$ = 1963 nm region. However, the Na and Al lines are beyond the range of this wavelength region. The $\lambda_C$ = 1947 nm region has the advantage of having a higher resolution than the $\lambda_C$ = 1963 nm region, resulting in more precise abundances. 

Altogether, the abundances of ten species can be available in the $\lambda_C$ = 1947 and 1963 nm windows (Table~\ref{stats}). This list of elements is similar to the list available from the $\lambda_C$ = 1180 nm region, with the addition of Na, Al, and O at the expense of K and Co. The elements in the K band would also allow us to probe the history and formation of the Milky Way, its stellar subsystems, and satellite galaxies in more detail than ever before because more stars in a much larger part of our Galaxy can be observed. O might be measured from the OH lines (which is a clear advantage over the $\lambda_C$ = 1180 nm region) and N from the CN lines, but this wavelength region, similarly to the $\lambda_C$ = 1180 nm region, contains no CO lines, so C might only be estimated using photometry. Even if C abundances might be determined with higher precision, this would increase the science return by measuring stellar ages and exploring the first dredge-up process in giant branch stars.

Excellent precision for radial velocities and atmospheric parameters for most FGKM stars is offered by selecting windows from the K band. With a Gaia-like setup and sampling of spectra on the detector, it is possible to achieve a precision of \vrad \ that is very similar to Gaia at a much longer wavelength. One disadvantage of the K band is the lack of absorption lines above 7000$-$8000~K, which severely limits the measurements of \vrad \ and atmospheric parameters for OBA-type stars, although the hydrogen Br$-\delta$ line offers a limited possibility to estimate the effective temperature and radial velocity. 

Larger window sizes would allow us higher resolutions and more precise measurements, but the crowding of spectra is expected to be more significant in the K band than it was for Gaia. This is especially true in the Milky Way disk and center. Thus, further studies are required to estimate the radial velocity and parameter precision at smaller window sizes, taking into account the issue of crowding when these simulations are available.

\section{Conclusions}
\label{sec4}

Our goal was to identify the optimal wavelength region between 800 and 2300 nm for the proposed GaiaNIR high-resolution spectrograph to maximize the precision of the derived radial velocities and atmospheric parameters. By generating 10\,000 synthetic spectra from the BOSZ library and cross-correlating these mock observations with ideal templates, we assessed the scatter of radial velocity residuals for resolutions ranging from 5000 to 20\,000. Based on also analyzing the estimated precision of atmospheric parameters in the selected candidate windows, our recommendations are listed below.

Primary recommendation: The wavelength window between 1926 and 2000 nm is our preferred strategic choice for the GaiaNIR spectrograph. There are three possibilities to use the multiple moderately high resolutions explored here: one possibility between 1926$-$2000 nm, and two narrower ranges at higher resolution: 1926$-$1968 nm and 1984$-$2000 nm. This last K-band region has the potential to achieve radial velocity precisions of 100 m/s or better for FGKM stars at the highest \snr, but it provides a relatively poor atmospheric parameter precision. The wide wavelength region gives a poorer radial velocity precision ($\approx$500 \ms) than the narrower windows. The optimal compromise between the radial velocity and atmospheric parameters might be achieved by observing between 1926$-$1968 nm with a maximum resolution of 20\,000 because its best radial velocity precision is about 160$-$260 \ms \ for FGKM stars and it also provides atmospheric parameter precisions better than 100~K for \teff and 0.08~dex for the metallicity. These values are highly competitive and on par with the current Gaia RVS capabilities. The interstellar reddening in the K band is lower by approximately 7.4 times than at optical wavelengths, so operating in this regime would make it possible to fulfill the GaiaNIR primary mandate of probing and mapping the dust-obscured regions of our Galaxy. The selected K-band windows will allow us to derive abundances for nine distinct species: O, Na, Mg, Al, Si, Ca, Ti, V, Cr, Mn, and Ni, depending on the chosen \teff, \logg, the wavelength region, and the resolution. However, only Ca can be measured from the spectra of M dwarfs due to strong H$_{\rm 2}$O blanketing. This can be a significant disadvantage as the majority of the GaiaNIR targets are expected to be red dwarfs. If the wider wavelength range is chosen (1926 $-$ 2000 nm), it would be highly desirable to increase the spectrograph resolution to about 14\,000 to further enhance the measurements of radial velocities and abundances, but crowding can become too much of an issue. Thus, further studies are required to assess the exact wavelength region and resolution before designing the spectrograph.

Secondary recommendation: We identified a secondary region centered around 1180 nm (1158 $-$ 1202 nm) that offers a wider effective temperature range (up to 12\,000$-$16\,000~K) for deriving parameters than the K band. However, because this region covers only slightly longer wavelengths than the Gaia optical spectrograph, it might fail to provide a substantial reduction in interstellar extinction, limiting the GaiaNIR utility for observing obscured Galactic regions. Our estimated \vrad \ precision in this region in the best-case scenario for FGKM stars is $\approx$ 340 \ms, which is about twice lower than the K band. The precision of the main atmospheric parameters is also somewhat lower than what is possible from the H band, but this is offset by the increased temperature range from which it is possible to derive them. The available species for abundance determination are the following: Mg, Si, K, Ca, Ti, V, Cr, Mn, Co, and Ni. One advantage of this wavelength window over the K band is that H$_{\rm 2}$O lines are not too strong in the spectra of M dwarfs. This not only facilitates normalizing their spectra (which is essential for accurate abundance measurements), but also facilitates determining the abundances of Mg, Si, K, Ca, Ti, and Cr for the coolest main-sequence stars below 4000~K instead of just Ca from the K band.

Ultimately, prioritizing the 1926$-$2000 nm window balances the need for high-precision radial velocities with the essential capability to observe the dust-obscured disk and far side of the Galaxy, ensuring GaiaNIR can successfully build upon the Gaia legacy. Significant effort will need to be put forward creating a new line list; this will be hard because the K band is heavily polluted by telluric lines in Earth-based observations. Future studies are needed to establish the achievable precision of all parameters and abundances to properly identify the optimal wavelength region and resolution between 1926 and 2000 nm for the GaiaNIR spectrograph. These future studies will need to address the issue of crowding. GaiaNIR is expected to observe significantly more stars than Gaia, and bright stars will contaminate the spectra of faint stars especially in the Galactic plane and the center of the Milky Way. The results from modeling the crowding will heavily affect the final selection of resolution, wavelength range, and the window size on the detector of the GaiaNIR spectrograph. Detailed studies of lower resolution options (allowing for more stars to be observed) are therefore necessary.

\begin{acknowledgements}
This project has been supported by the LP2021-9 Lend\"ulet grant of the Hungarian Academy of Sciences. On behalf of the "Finding the optimal wavelength range for the GaiaNIR spectrograph" project we are grateful for the possibility to use HUN-REN Cloud (see \citealt{heder2022}; https://science-cloud.hu/) which helped us achieve the results published in this paper. 
\end{acknowledgements}

\bibliographystyle{aa}
\bibliography{references}

\begin{appendix}
\onecolumn
\section{Additional figures}

This appendix shows various figures that show the difference of the fitted and original radial velocities and atmospheric parameters as a function of effective temperature. These figures help understanding in detail the effective temperature region in which reliable parameters can be derived. For detailed discussion of these figures see Sec.~\ref{sec3}.

\begin{figure}[ht]       
\centering
\includegraphics[width=5.45in,angle=0]{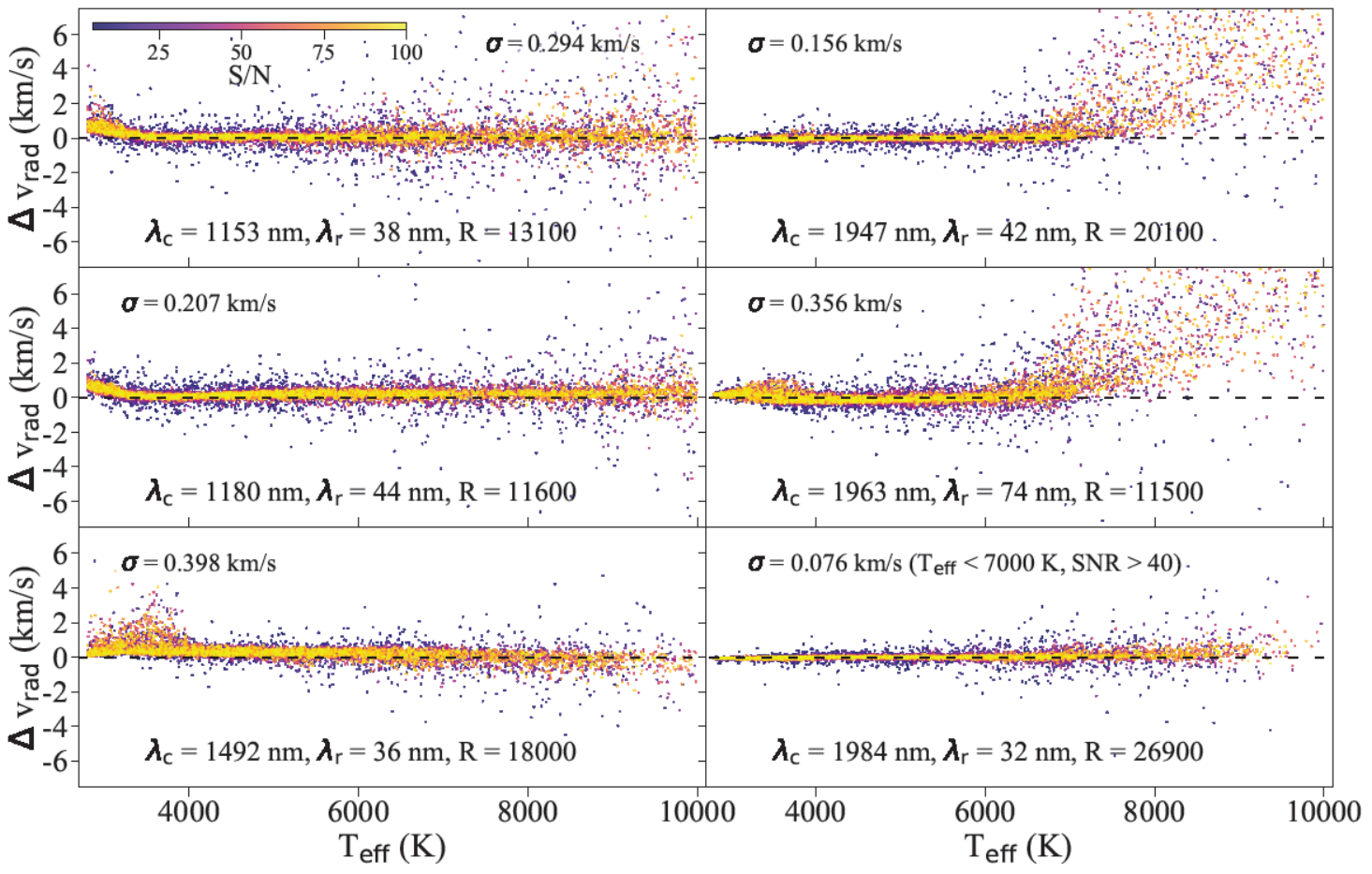}
\caption{Radial velocity differences as a function of \teff \ color coded by S/N for the six wavelength windows selected in Section 2. The relative lack of points in the $\lambda_{\rm C}$ = 1492 and 1984 nm panels are due to the fact that most of the differences being larger than the range of the plot. This indicates poor measurements because at high temperatures the stellar spectra contain only very few and weak absorption lines.}
\label{radveldiff}
\end{figure}

\begin{figure}[ht]
\centering
\includegraphics[width=5.45in,angle=0]{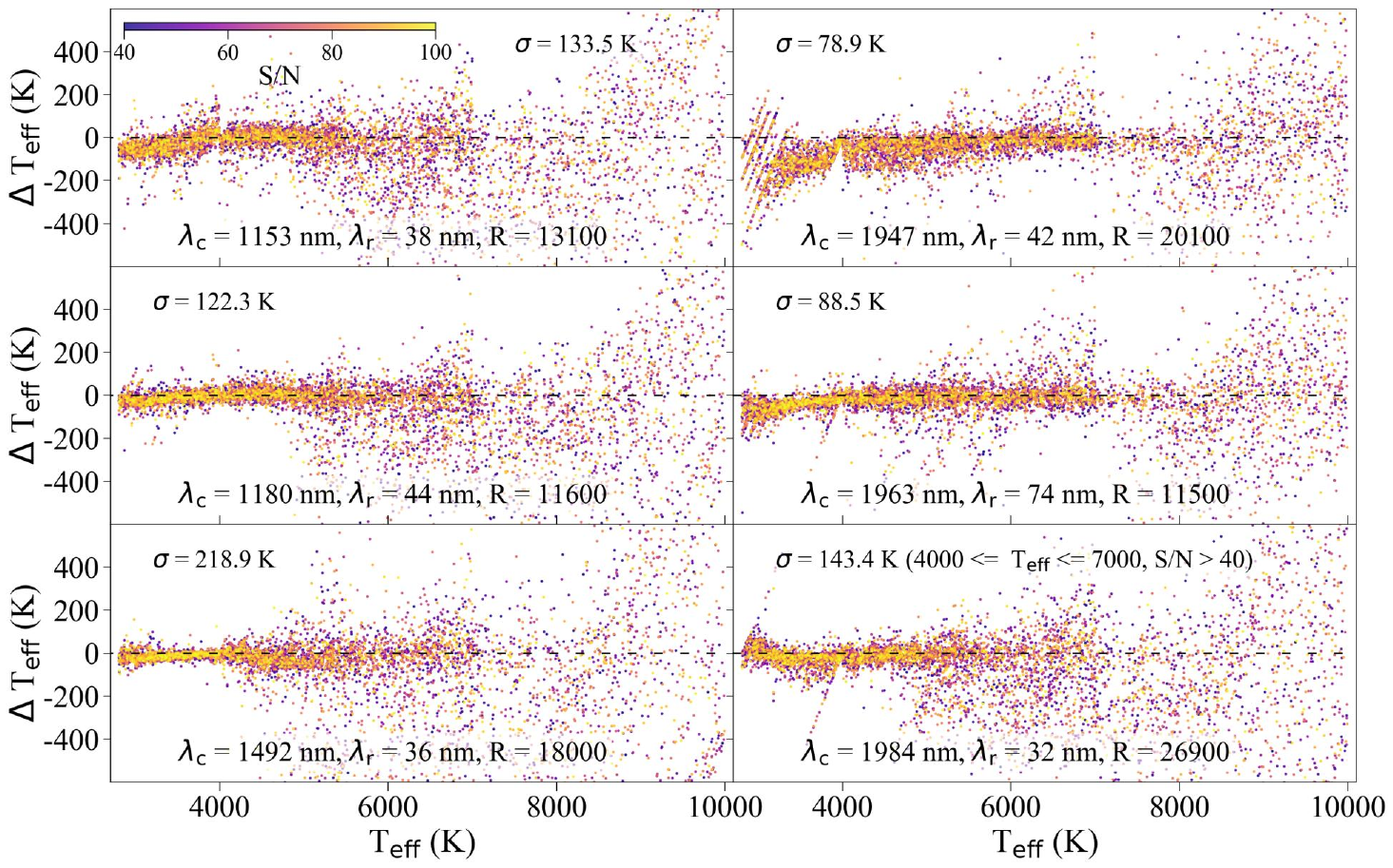}
\caption{Differences of \teff \ values as a function of the original \teff \ color coded by S/N. Atmospheric parameters are reliable at higher S/N values, so only tests with S/N $>$ 40 are plotted. The diagonal straight lines seen in some of the panels are noding effects arising when the spectra are very weakly sensitive to the particular parameter, which leads to FERRE finding the grid parameters as the best fit.}
\label{apcomp1}
\end{figure}

\FloatBarrier

\begin{figure}[ht]              
\centering
\includegraphics[width=5.6in,angle=0]{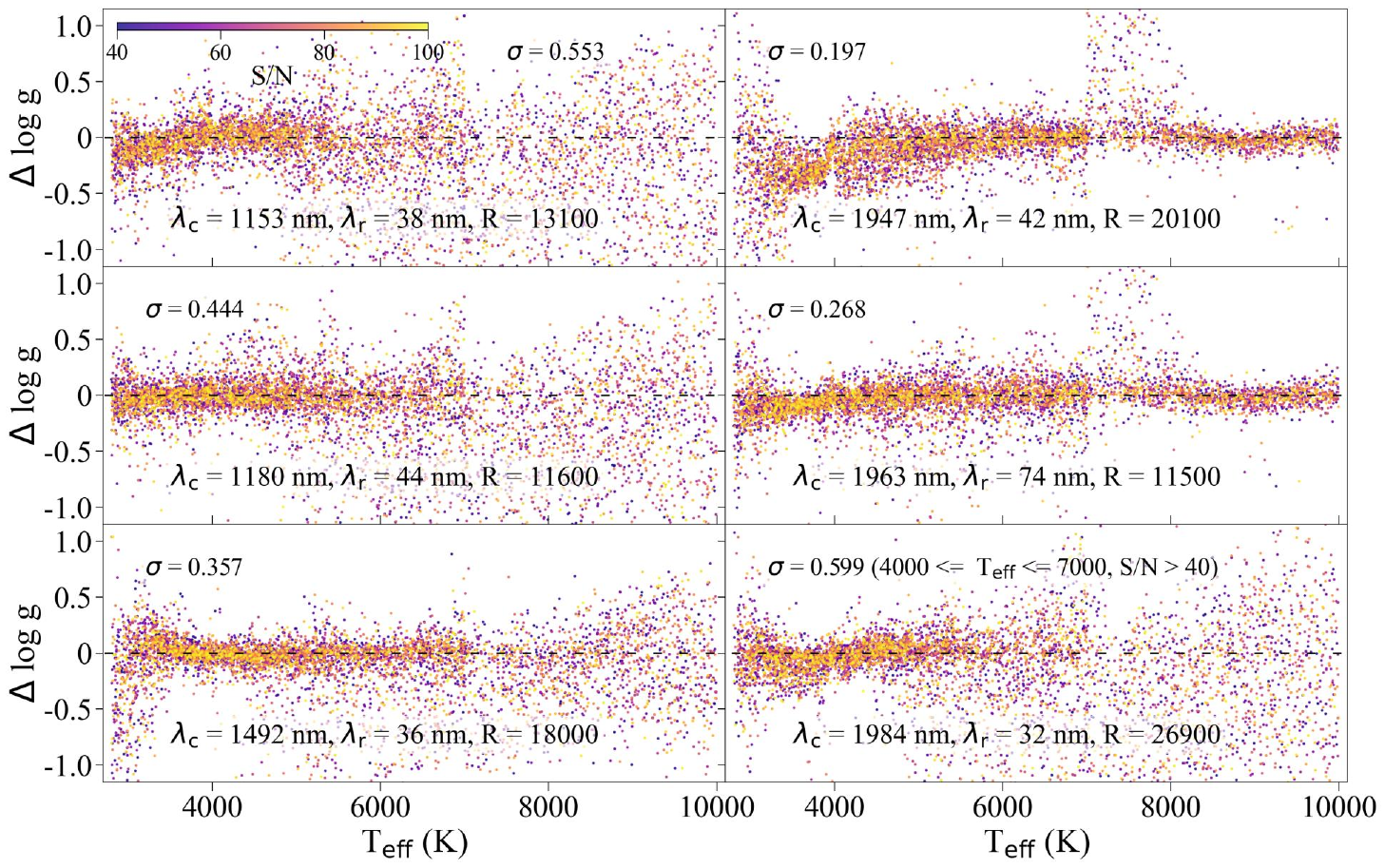}
\caption{Differences of \logg \ values as a function of the original \teff \ color coded by S/N. Atmospheric parameters are reliable at higher S/N values, so only tests with S/N $>$ 40 are plotted. We see large scatter of \logg \ differences because spectral lines are only weakly sensitive to changes of surface gravity.}
\label{apcomp2}
\end{figure}

\begin{figure}[ht]          
\centering
\includegraphics[width=5.6in,angle=0]{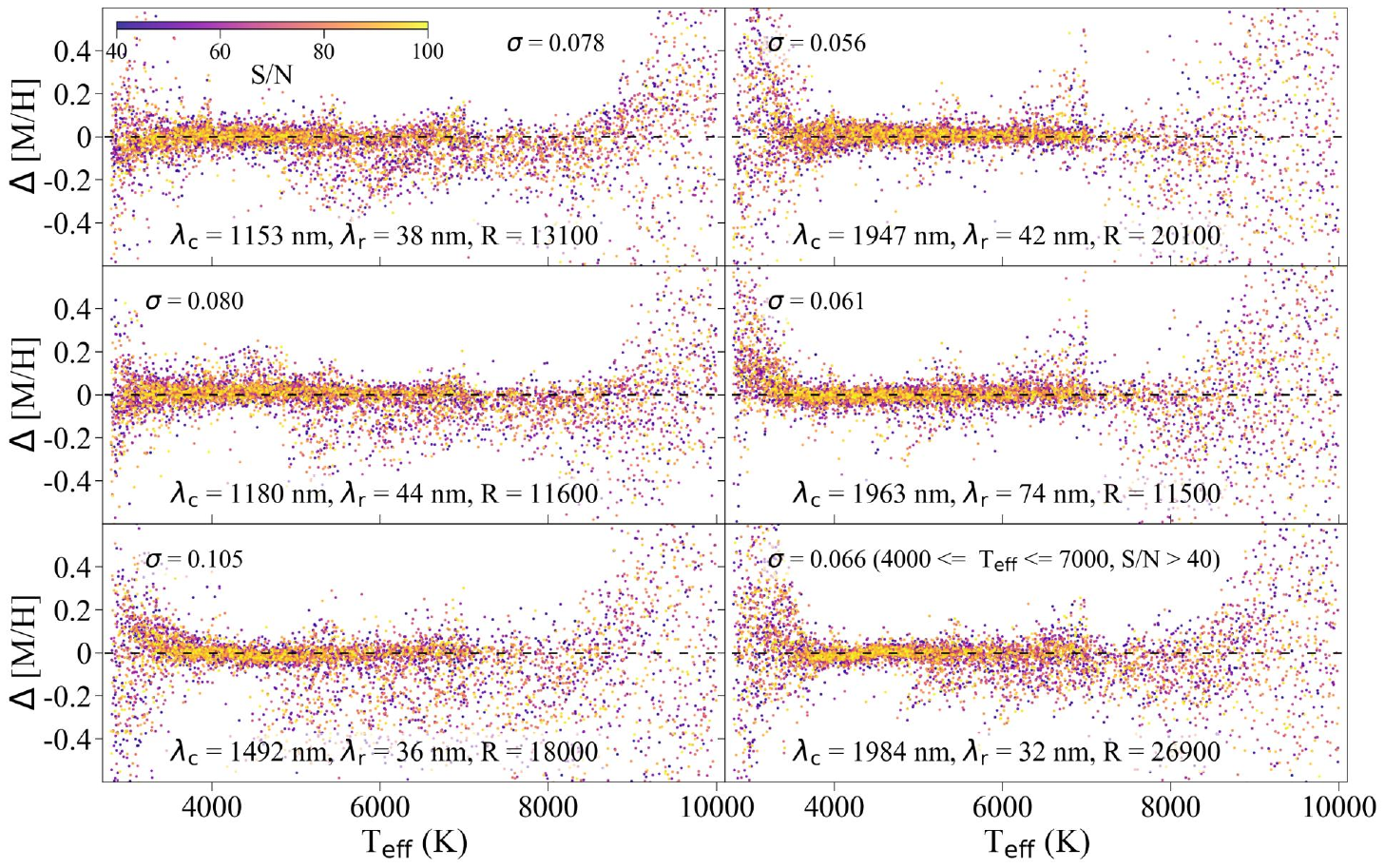}
\caption{Differences of \mh \ values as a function of the original \teff \ color coded by S/N. Atmospheric parameters are reliable at higher S/N values, so only tests with S/N $>$ 40 are plotted. Notice how difficult it becomes to measure metallicity reliably below \teff \ $<$ 3800~K from most windows, except in the case of $\lambda_{\rm C}$ = 1180 nm.}
\label{apcomp3}
\end{figure}

\end{appendix}

\end{document}